\documentclass{article}
\usepackage{amsmath}
\usepackage{amssymb}
\usepackage{amsthm}
\usepackage{amsbsy}
\usepackage{graphicx}
\usepackage{cite}

\newcommand{\abs}[1]{\left|#1\right|}
\newcommand{\atpoint}[2]{\left.#1\right|_{#2}}
\newcommand{\FU}{\mathfrak{U}}
\newcommand{\FV}{\mathfrak{V}}
\newcommand{\frb}{\mathfrak{b}}
\newcommand{\Pp}{\mathcal{P}}
\newcommand{\Uu}{\mathcal{U}}

\newcommand{\ol}{\lambda^{\ast}}
\newcommand{\bheta}{\boldsymbol{\eta}}
\newcommand{\bzeta}{\boldsymbol{\zeta}}
\newcommand{\bzero}{\boldsymbol{0}}
\newcommand{\bD}{\boldsymbol{\Delta}}
\newcommand{\ba}{\mathbf{a}}
\newcommand{\bb}{\mathbf{b}}
\newcommand{\be}{\mathbf{e}}
\newcommand{\bm}{\mathbf{m}}
\newcommand{\bn}{\mathbf{n}}
\newcommand{\bk}{\mathbf{k}}
\newcommand{\bu}{\mathbf{u}}
\newcommand{\bU}{\mathbf{U}}
\newcommand{\bV}{\mathbf{V}}

\newcommand{\bZ}{\mathbf{Z}}
\newcommand{\bbR}{\mathbb{R}}
\newcommand{\bbZ}{\mathbb{Z}}
\newcommand{\bbC}{\mathbb{C}}
\newcommand{\whs}{\widehat{\sigma}}
\newcommand{\wht}{\widehat{\tau}}
\newcommand{\wkk}{\widetilde{k}}
\newcommand{\wth}{\widetilde{h}}
\newcommand{\wtk}{\widetilde{\kappa}}

\newcommand{\vphi}{\varphi}
\newcommand{\dn}{\mathrm{dn}}
\newcommand{\spb}{St.-Petersburg State University of Aerospace Instrumentation (SUAI),
67 Bolshaya Morskaya str., St.-Petersburg, 190000,Russia}
\newcommand{\alaska}{University of Alaska Fairbanks, 505 South Chandalar Drive,
Fairbanks, AK 99775, USA}

\newtheorem{thm}{Theorem}

\allowdisplaybreaks

\begin{document}

\title{On a periodic solution of the focusing nonlinear Schr\"odinger equation}

\author{A.O.~Smirnov\footnote{Corresponding author: alsmir@guap.ru} {}\footnote{\spb}, E.G.~Semenova \addtocounter{footnote}{-1}\footnotemark, V.~Zinger \addtocounter{footnote}{-1}\footnotemark { }\footnote{\alaska}, and N.~Zinger \addtocounter{footnote}{-2}\footnotemark { }\footnotemark}

\date{}

\maketitle

\begin{abstract}
A periodic two-phase algebro-geometric solution of the focusing nonlinear Schr\"odinger equation is constructed in terms of elliptic Jacobi theta-functions.
A dependence of this solution on the parameters of a spectral curve is investigated.
An existence of a real smooth finite-gap solution of NLS equation with complex initial phase is proven.
Degenerations of the constructed solution to one-phase traveling wave solution and solutions in the form of the plane waves are carried.
\end{abstract}


MSC (2010) 35Q55, 37C55

\section*{Introduction}

Currently, one of the most hot topic of nonlinear physics and mathematics is an investigation of the nature and prediction of appearance so called ``rogue waves'' or ``freak waves''. Freak waves are very popular in last time solutions that represent local short-term grows of amplitude or  ``waves that appear from nowhere and disappear without a trace'' \cite{AAT}.
Although freak waves were detected in models concerning various fields of physics (see for example \cite{EPJ}) main fields of they appearance are hydrodynamics \cite{Per83, Zakh08, Chab11, Chab12} and nonlinear optics \cite{AkhAnke, Kib10, Kib12}.

In general, a study of freak waves in principle approximation is based on a consideration of the focusing nonlinear Schr\"odinger equation (NLS)
\begin{equation}
ip_t+p_{xx}+2\abs{p}^2p=0, \quad i^2=-1. \label{eq:nls}
\end{equation}
Last time physicians and mathematicians studied very actively rational solutions of \eqref{eq:nls}; usually these solutions are obtained by Darboux transformation and its generalizations \cite{DGKM10, DubMat11, AKA, AKA2, Ohta12, HeF, GenDT11, DubMat13, GP13, GP14}. Exact solutions distinguished from rational ones are mentioned quite seldom \cite{Aek85e,AK85e, IAKe, IKe, Chow95, Chow02, Osb00, AkhAnke, Sch02, Sch06, Kib12}.

While taking into account more fine effects \cite{Zakh08, AkhAnke, Dys79, Dys96, Dys00, Ind09, Sch05, Sch06} equations under consideration differ from \eqref{eq:nls} by usage of the additional terms leading to non-integrability of an equation. Therefore, a construction of some classes of solutions becomes impossible.  At the same time there exist such situations when a mentioned additional terms may be omitted \cite{AkhAnke, Sch05, Sch06}. In this case a consideration integrable equation \eqref{eq:nls} allows to use solution classes intrinsic to integrable equations such as multi-phase periodic solutions.

On the one hand, several works \cite{Kib10, Chab11, Chab12} contain statement that in the laboratory experiments some waves were observed with the characteristics close to those of rational solutions. On the other hand, the outcomes of the studies \cite{Sch05, Sch06} show that observed in experiments waves are close to multi-phase solutions. However, there is a possibility that if authors of studies \cite{Kib10, Chab11, Chab12} would compare experimental data with multi-phased solutions' characteristics, then they would obtain relatively close agreement. Especially it may be so in the cases when the sizes of cell of lattice of periods are considerably larger than dimensions of tank or  waveguide.

Of course, one can not expect the strictly periodic rogue waves to appear on a water surface. However, in case of nonlinear optics, is quite possible to create situations leading to an appearance of periodic freak waves.  For example, in the research \cite{Kib12} authors discuss an observation Kuznetsov-Ma soliton  \cite{Ku77e, Ma79} in the experiment. Let us remark that this solution is a limit case of two-phase algebro-geometric solution \cite{Sm12pomie}. Moreover, usage appropriate limit in multi-phase solutions \cite{BBEIM, Sm12pomie, Kalla11, Kalla11a} allows us to obtain a rational, solitonic or periodic in $x$ homoclinic solutions \cite{Sch05, Sch06}, i.e., solutions that are usually obtained with the help of Darboux transformations. It would be interesting to know what multi-phase solutions lead to symmetrical rational solutions from \cite{AKA, AKA2}.

In the present paper we consider two-phase periodic in $x$ and in $t$ algebro-geometric solution of equation \eqref{eq:nls}. Considered two-phase solution is a particular case of multi-phase solutions.
Multi-phase algebro-geometric solutions are constructed by ``finite-gap (algebro-geometric) integration method'' \cite{BBEIM, Soleq, Soleq2}; this method was created in the works of Dubrovin, Novikov, Marchenko, Lax, McKean, van Moerbeke, Matveev, Its, Krichever \cite{Nov74e, Lax74, DNe, Mar74e, MvM, IMe, DMNe, Kr77e, Dub81e}. One who is interested in details of the development of the method is to refer to the review article \cite{Mat08}. It should be mentioned that another method of constructing multi-phase algebro-geometric solutions of integrable nonlinear equations exists \cite{MumII,Prev85, Kalla11, Kalla11a, KK12, KK12a}. Let us remark that first method is based on Baker-Akhiezer function but second method is based on some Fay's identities \cite{Fay}. In our paper we use first method and Its' and Kotlyarov's classic formulas \cite{Its76e, IKe} (see also \cite{BBEIM}).

Our aim here is to understand an influence of parameters of a spectral curve on the behavior and the shape of two-phase algebro-geometric solutions of \eqref{eq:nls}.  The first section of this paper contains the basic notations and classic formulas of algebro-geometric solutions of the focusing nonlinear Schr\"odinger equation \eqref{eq:nls}.
The second section of the paper is devoted to the construction of an example of two-phase algebro-geometric solution; As in \cite{Sm12tmfe, Sm13mze}, new solution is expressed in Jacobi elliptic theta functions. It should be noted that present solution differ from previous theta-functional solutions \cite{Chow95, Chow02, Sm12tmfe, Sm13mze}. In particular, in present paper all branch points have one same real parts, while in \cite{Sm12tmfe, Sm13mze} all pairs of branch points have different real parts. As we can see, this leads to the following results. Crests of amplitude of constructed periodic solution disposed in the nodes of a quadrangular lattice. Crests of previous solutions formed two quadrangular lattice; nodes of one lattice disposed in centers of parallelograms of second lattice. Therefore, considered in present paper solution is a special case of two-phase solutions.
The third section contains an analysis of dependence of solution on the parameters of a spectral curve.
In particular, we show that if a distance between branching points decreases, then a steepness of front of solution increases. Also in present work we consider for the first time a dependence of two-phase solution of equation \eqref{eq:nls} from a sum of branch points. We show that if a sum of branch points differs from zero then then solution of equation \eqref{eq:nls} has not rectangular lattice of periods, and if sum equals zero then lattice is rectangular. Another new statement is following: a real smooth finite-gap solution of NLS equation may have a complex initial phase. In \cite{BBEIM} one can found a statement about necessity of a real initial phase for smooth real finite-gap solution.
In the fourth section we calculate some limits of the investigated two-phase solution. These limits include a new theta-functional expression for one-phase solution. Some technical details of related calculations can be found in Appendix.

\section{Finite-gap solutions of the focusing nonlinear Schr\"odinger equation}

We use well-known method of construction finite-gap solutions of NLS equation \cite{IKe, BBEIM, Soleq, Sm94msbe, Sm96tmfe}. It is based on the fact that the equation \eqref{eq:nls} can be obtained from the coupled nonlinear Schr\"odinger equation (cNLS)
\begin{equation}
\begin{cases}
i p_t+p_{xx}-2p^2q= 0,\\
i q_t-q_{xx}+2pq^2= 0
\end{cases} \label{eq:nls.cases}
\end{equation}
by reduction $q=-p^{\ast}$.  In turn, the cNLS \eqref{eq:nls.cases} results from the compatibility condition of the following system of linear equations (Lax pair)
\begin{gather}
\Psi_x=\FU \Psi, \label{op:dir}\\
\Psi_t=\FV \Psi,
\end{gather}
where
\begin{gather*}
\FU=\lambda\begin{pmatrix} -i & 0\\ 0 & i \end{pmatrix}+\begin{pmatrix} 0 & ip\\ -iq & 0 \end{pmatrix},\quad
\FV=2\lambda\FU+\begin{pmatrix}-ipq&-p_x\\ -q_x& ipq\end{pmatrix}.
\end{gather*}

Finite-gap solutions of system \eqref{eq:nls.cases} are parameterized by the hyper-elliptic curve
$\Gamma=\{(w,\lambda)\}$ of the genus~$g$:
\begin{equation*}
\Gamma:\quad w^2= \prod_{j=1}^{2g+2}(\lambda-\lambda_j).
\end{equation*}
The branch points ($\lambda= \lambda_j$, $j=1,\ldots,2g+2$) of this curve are the endpoints of the spectral arcs of continuous spectrum of Dirac operator \eqref{op:dir}. Infinitely far point of the spectrum corresponds to two different  points $\Pp_{\infty}^{\pm}$  on the curve $\Gamma$.
If $q=-p^\ast$, then the curve $\Gamma$ has the form
\begin{equation}
\Gamma:\quad w^2=\prod_{j=1}^{g+1}(\lambda-\lambda_j)(\lambda-\ol_j),\quad \Im \lambda_j\ne 0. \label{nls:curve}
\end{equation}

Following a standard procedure of constructing a finite-gap solutions \cite{BBEIM, Dub81e}, let us to choose on $\Gamma$ a canonical basis of cycles $\gamma^t=(a_1,\ldots,a_g,b_1,\ldots,b_g)$ with matrix of indices of intersection
\begin{equation*}
C_0=\begin{pmatrix} 0&I\\-I&0 \end{pmatrix}.
\end{equation*}
The condition $q=-p^{\ast}$ implies \cite{BBEIM, Dub81e} that the basis of cycles should satisfy the relation:
\begin{equation}
\wht_1 \ba=-\ba,\quad \wht_1 \bb=\bb+K\ba, \label{wht1}
\end{equation}
where $\tau_1$ is an anti-holomorphic involution
\begin{equation*}
\tau_1:(w,\lambda)\to(w^{\ast},\lambda^{\ast}).
\end{equation*}
A normalized basis of holomorphic differentials $d\Uu_j$ corresponds to
the canonical basis of cycles
\begin{equation} \label{cond:norm}
\oint_{a_k}d\Uu_j=\delta_{kj},\quad k,j=1,\ldots,g.
\end{equation}
It is well known \cite{Bake, Spre, FKra, Dub81e} that a matrix of periods $B$ of a curve $\Gamma$,
\begin{equation} \label{def:B}
B_{kj}=\oint_{b_k}d\Uu_j,\quad k,j=1,\ldots,g,
\end{equation}
is a symmetric matrix with positively defined imaginary part.

Let us define $g$-dimensional Riemann theta function with characteristics $\bheta,\bzeta\in\bbR^g$ \cite{Fay, Bake, Dub81e}:
\begin{equation} \label{def:theta}
\begin{gathered}
\Theta[\bheta^t;\bzeta^t](\bu|B)=\sum_{\bm\in\bbZ^g}\exp\{\pi i (\bm+\bheta)^tB(\bm+\bheta)+2\pi i(\bm+\bheta)^t(\bu+\bzeta)\},\\
\Theta[\bzero^t;\bzero^t](\bu|B)\equiv\Theta(\bu|B),
\end{gathered}
\end{equation}
where $\bu\in\bbC^g$, and summation pass over $g$-dimensional integer lattice.

Let us also define on $\Gamma$ normalized Abelian integrals of a second kind $\Omega_1(\Pp)$, $\Omega_2(\Pp)$ and third kind $\omega_0(\Pp)$  with the following asymptotic at infinitely far points $\Pp_{\infty}^{\pm}$:
\begin{align*}
&\oint_{a_k}d\Omega_1
=\oint_{a_k}d\Omega_2
=\oint_{a_k}d\omega_0= 0,
&&k=1,\ldots,g, \\
&\Omega_1(\Pp)= \mp i \left(\lambda+K_1
+O\left(\lambda^{-1}\,\right)\right),
&&\Pp\to\Pp_{\infty}^{\pm}, \\
&\Omega_2(\Pp)= \mp i \left(2\lambda^2+K_2
+O\left(\lambda^{-1}\,\right)\right),
&&\Pp\to\Pp_{\infty}^{\pm}, \\
&\omega_0(\Pp)= \mp\left(\ln \lambda-\ln K_0
+O\left(\lambda^{-1}\,\right)\right),
&&\Pp\to\Pp_{\infty}^{\pm}, \\
& w = \pm\left( \lambda^{g+1}+O\left(\lambda^g\,\right)\right),
&&\Pp\to\Pp_{\infty}^{\pm}.
\end{align*}

Let us denote by  $2\pi i \bU $, $2\pi i \bV $
the vectors of $b$-periods of Abelian integrals of the second kind
$\Omega_1(\Pp)$, $\Omega_2(\Pp)$ respectively.

\begin{thm}[\cite{IKe}] Solution of cNLS \eqref{eq:nls.cases} has a form
\begin{equation}
\begin{aligned}
&p(x,t)= \frac{2K_0}{A}\frac{\Theta(\bZ)\Theta( \bU x+ \bV t+\bZ-\bD)}%
{\Theta(\bZ-\bD)\Theta( \bU x+ \bV t+\bZ)}\exp\{2 i \Phi(x,t)\},\\
&q(x,t)= 2A K_0\frac{\Theta(\bZ-\bD)\Theta( \bU x+ \bV t+\bZ+\bD)}%
{\Theta(\bZ)\Theta( \bU x+ \bV t+\bZ)}\exp\{-2 i \Phi(x,t)\},\\
\end{aligned}\label{eq:p,q}
\end{equation}
where $\Phi(x,t)= K_1 x+K_2 t$.
Vector $\bD$ is a vector of Abelian holomorphic integrals that are calculated along the path between points $\Pp_{\infty}^{-}$ and $\Pp_{\infty}^{+}$, and this path does not intersect any of the basic cycles. The $\bZ$ is initial phase of solution, $A\ne0$ is an arbitrary constant.
\end{thm}

Eqs.  \eqref{eq:p,q}, \eqref{nls:curve} imply that for the focusing NLS equation an amplitude of solution $\abs{p}$ satisfies the equality
\begin{equation}
\abs{p}^2=-4K_0^2\dfrac{\Theta(\bU x+\bV t+\bZ-\bD) \Theta(\bU x+\bV t+\bZ+\bD)}
{\Theta^2(\bU x+\bV t+\bZ)}, \label{eq:abs.p}
\end{equation}
where
\begin{equation*}
\Im\bU=\Im\bV=\Im\bZ=\bzero,\quad K_0^2<0.
\end{equation*}

\section{The curve of genus $g=2$ with involution}

\begin{figure}[htb]
\begin{center}
\includegraphics[width=0.45\textwidth]{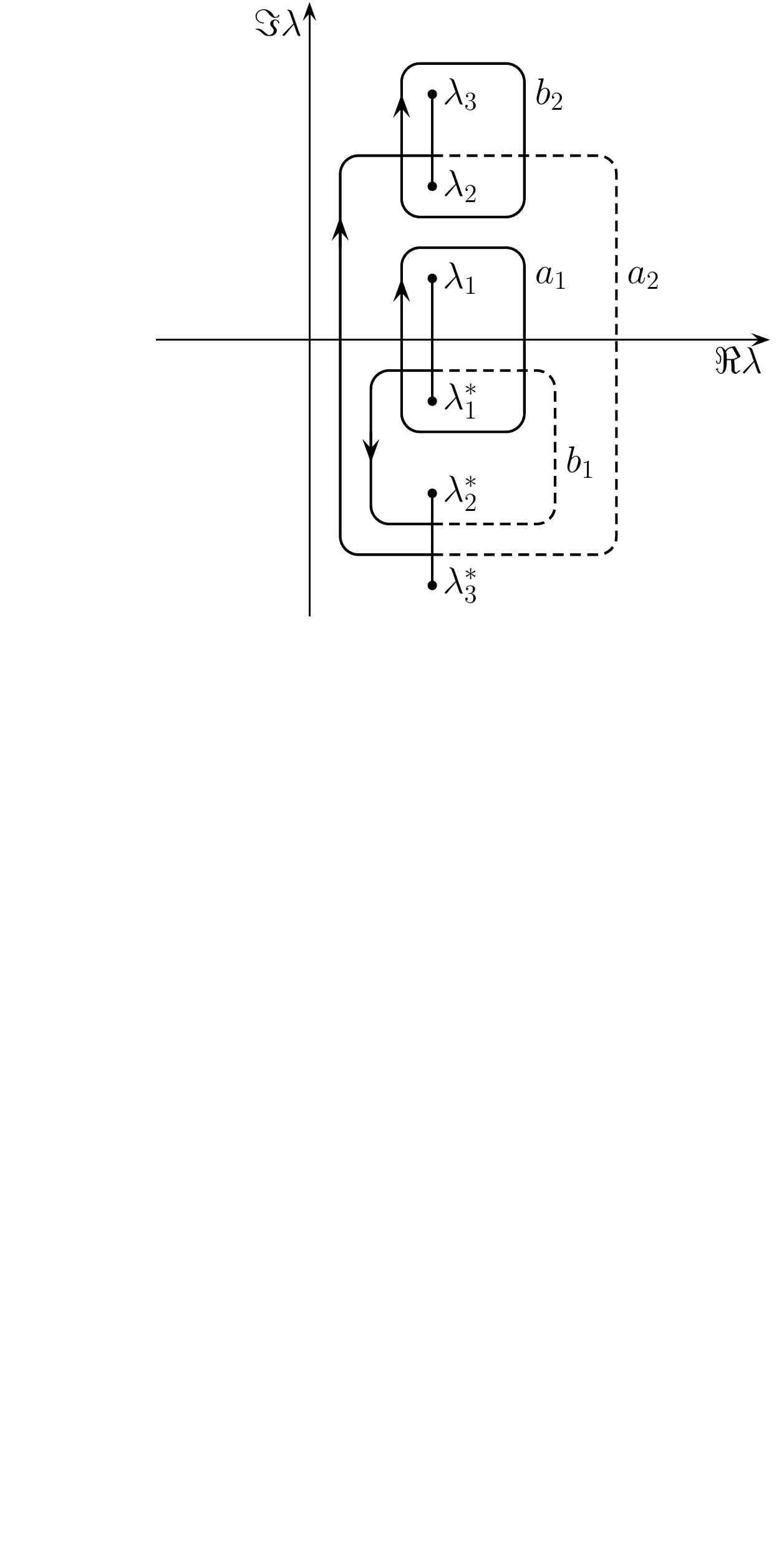}
\caption{The curve $\Gamma_{2}$}\label{fig:gamma2}
\end{center}
\end{figure}

To construct an example of solution we will use the curve $\Gamma_2$ (fig.~\ref{fig:gamma2}); this curve is defined by equation
\begin{equation}
w^2=(\lambda^2-2\lambda_0\lambda+|\lambda_1|^2) (\lambda^2-2\lambda_0\lambda+|\lambda_2|^2) (\lambda^2-2\lambda_0\lambda+|\lambda_3|^2), \label{eq:gamma2}
\end{equation}
where
\begin{equation*}
\Re\lambda_1=\Re\lambda_2=\Re\lambda_3=\lambda_0,\quad
0<\Im\lambda_1<\Im\lambda_2<\Im\lambda_3.
\end{equation*}

Let us select the basis of cycles $\Gamma_2$ as on the fig.~\ref{fig:gamma2} and the basis of normalized holomorphic differentials:
\begin{equation*}
d\Uu_j=\dfrac{(c_{j1}\lambda+c_{j2})d\lambda}{w}.
\end{equation*}.

\begin{figure}[bt]
\begin{center}
\parbox{0.45\textwidth}{\includegraphics[width=0.45\textwidth]{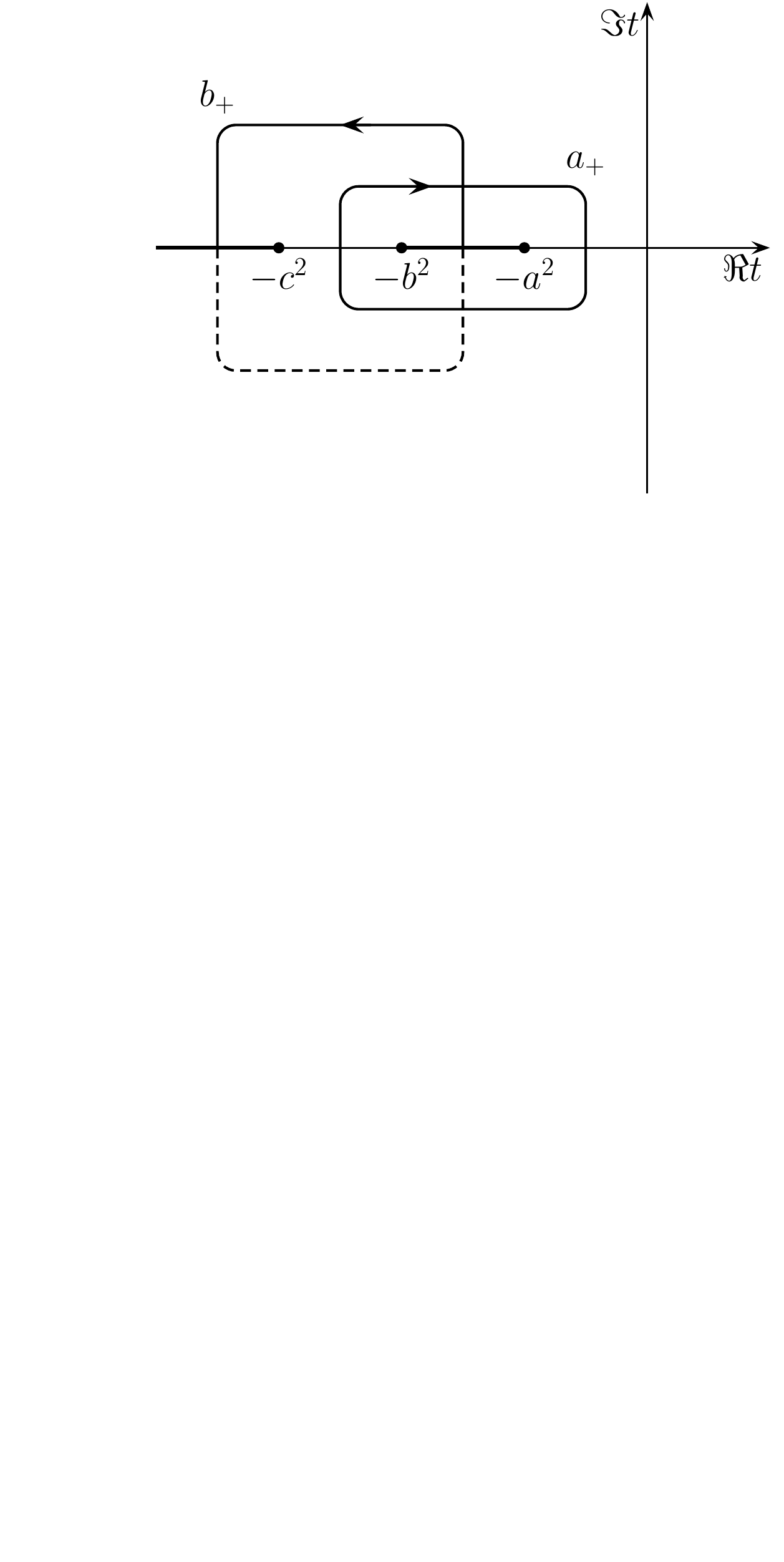}
\caption{The curve $\Gamma_{+}$} \label{fig:Gamma+}}\hfil
\parbox{0.45\textwidth}{\includegraphics[width=0.45\textwidth]{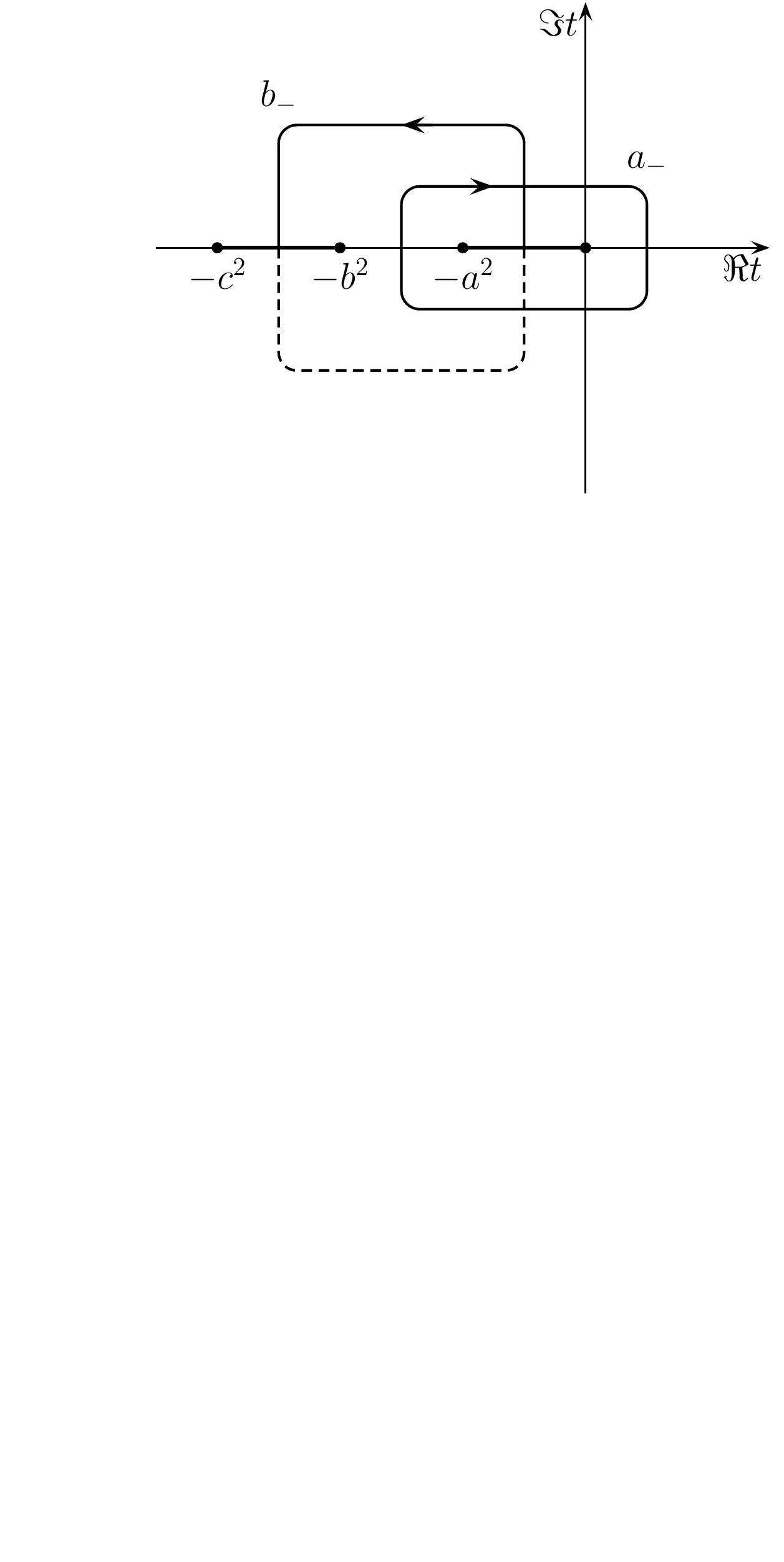}
\caption{The curve $\Gamma_{-}$} \label{fig:Gamma-}}
\end{center}
\end{figure}

Since two holomorphic involutions exist on $\Gamma_2$:
\begin{align}
&\tau_0:(w,\lambda)\to(-w,\lambda),\\
&\tau_2:(w,\lambda)\to(w,2\lambda_0-\lambda),
\end{align}
then the curve $\Gamma_2$ covers two elliptic curves: $\Gamma_{+}=\Gamma/\tau_2$  (fig.~\ref{fig:Gamma+})
\begin{equation}
\Gamma_{+}:\quad \chi_{+}^2=(t+a^2)(t+b^2)(t+c^2)
\end{equation}
and $\Gamma_{-}=\Gamma/(\tau_0\tau_2)$ (fig.~\ref{fig:Gamma-})
\begin{equation}
\Gamma_{-}:\quad \chi_{-}^2=t(t+a^2)(t+b^2)(t+c^2),
\end{equation}
where
\begin{equation*}
a=\Im\lambda_1,\quad b=\Im\lambda_2,\quad c=\Im\lambda_3.
\end{equation*}

The covering mappings are given by the formulae
\begin{gather}
t=(\lambda-\lambda_0)^2,\quad \chi_{+}=w,\quad \chi_{-}=(\lambda-\lambda_0)w, \label{def:cov.t}\\
\dfrac{dt}{\chi_{+}}=\dfrac{2(\lambda-\lambda_0) d\lambda}{w},\quad
\dfrac{dt}{\chi_{-}}=\dfrac{2 d\lambda}{w}. \label{def:cov.du}
\end{gather}

The existence of covering lead \cite{Sm87msbe, Sm89tmfe} to following proposition.

\begin{thm}
Generated by the spectral curve $\Gamma_2$ two-phase solution can be expressed by elliptic theta functions.
Parameters of related solution can be expressed in terms of elliptic integrals on the curves  $\Gamma_{\pm}$.
\end{thm}

It follows from equalities \eqref{def:cov.t} that covering mappings generate the following transformations between bases of cycles
\begin{equation}
\begin{aligned}
&\whs_{+}\binom{a_1}{a_2}=\begin{pmatrix}0&0\\-2&0\end{pmatrix}\binom{a_{+}}{b_{+}}, \quad
&&\whs_{-}\binom{a_1}{a_2}=\begin{pmatrix}2&0\\0&0\end{pmatrix}\binom{a_{-}}{b_{-}}, \\
&\whs_{+}\binom{b_1}{b_2}=\begin{pmatrix}1&0\\0&-1\end{pmatrix}\binom{a_{+}}{b_{+}}, \quad
&&\whs_{-}\binom{b_1}{b_2}=\begin{pmatrix}0&1\\-1&0\end{pmatrix}\binom{a_{-}}{b_{-}}.
\end{aligned} \label{eq:cov.cycl}
\end{equation}

It follows from relations \eqref{def:cov.du}, \eqref{eq:cov.cycl} that the matrix of coefficients of normalized holomorphic differentials $d\Uu_j$ equals to
\begin{equation*}
C=\begin{pmatrix}
0& i/(2A_{-})\\ -i/(2A_{+})& i\lambda_0/(2A_{+})
\end{pmatrix},
\end{equation*}
where
\begin{equation*}
A_{+}=\int_{a^2}^{b^2}\dfrac{dt}{\sqrt{(t-a^2)(b^2-t)(c^2-t)}},\quad
A_{-}=\int_0^{a^2}\dfrac{dt}{\sqrt{t(a^2-t)(b^2-t)(c^2-t)}}.
\end{equation*}

Calculating the matrix of the periods of the curve $\Gamma_2$ we have
\begin{equation*}
B=\begin{pmatrix}
i\frb_{-}/2& -1/2\\ -1/2& i\frb_{+}/2
\end{pmatrix},
\end{equation*}
where $\frb_{\pm}=B_{\pm}/A_{\pm}$,
\begin{equation*}
B_{+}=\int_{b^2}^{c^2}\dfrac{dt}{\sqrt{(t-a^2)(t-b^2)(c^2-t)}},\quad
B_{-}=\int_{a^2}^{b^2}\dfrac{dt}{\sqrt{t(t-a^2)(b^2-t)(c^2-t)}}.
\end{equation*}

It follows from Riemann bilinear relations \cite{Dub81e, Bake} that $b$-periods of the normalized Abelian differentials of the second kind are equal
\begin{gather*}
U_m=-i\left(\atpoint{\dfrac{d\Uu_m}{d\xi_{-}}}{\xi_{-}=0} -\atpoint{\dfrac{d\Uu_m}{d\xi_{+}}}{\xi_{+}=0}\right)=-2ic_{m1},\\
V_m=-2i\left(\atpoint{\dfrac{d^2\Uu_m}{d\xi_{-}^2}}{\xi_{-}=0} -\atpoint{\dfrac{d^2\Uu_m}{d\xi_{+}^2}}{\xi_{+}=0}\right)=-12i\lambda_0c_{m1}-4ic_{m2},
\end{gather*}
where $\xi_{\pm}=\lambda^{-1}$ are local parameters in the neighborhood of the infinitely far points  $\Pp_{\infty}^{\pm}$, and $c_{mj}$ are coefficients of holomorphic differentials $d\Uu_m$.

Calculating derivatives we get
\begin{equation*}
U_1=0,\quad U_2=-1/A_{+},\quad V_1=2/A_{-}, \quad V_2=-4\lambda_0/A_{+}.
\end{equation*}

Changing in formula \eqref{def:theta} a summation with respect to $\bm$ on a summation with respect to $\bn$ and $\bk$:
\begin{equation*}
m_j=2n_j+k_j,\quad n_j\in\bZ,\quad k_j\in\{0;1\},\quad j=1,2
\end{equation*}
we obtain
\begin{multline*}
\Theta(\bu|B)=\vartheta_3(2u_1|2i\frb_{-})\vartheta_3(2u_2|2i\frb_{+})+
\vartheta_2(2u_1|2i\frb_{-})\vartheta_3(2u_2|2i\frb_{+})+\\
+\vartheta_3(2u_1|2i\frb_{-})\vartheta_2(2u_2|2i\frb_{+})-
\vartheta_2(2u_1|2i\frb_{-})\vartheta_2(2u_2|2i\frb_{+}),
\end{multline*}
where $\vartheta_j$ are Jacoby elliptic theta functions \cite{Akhe}
\begin{align*}
&\vartheta_1(u|b)=2\sum_{m=1}^\infty (-1)^{m-1}h^{(m-1/2)^2}\sin[(2m-1)\pi u],\quad h=e^{\pi i b},\\
&\vartheta_2(u|b)=2\sum_{m=1}^\infty h^{(m-1/2)^2}\cos[(2m-1)\pi u],\\
&\vartheta_3(u|b)=1+2\sum_{m=1}^\infty h^{m^2}\cos(2m\pi u),\\
&\vartheta_4(u|b)=1+2\sum_{m=1}^\infty (-1)^m h^{m^2}\cos(2m\pi u).
\end{align*}

Therefore, associated with curve \eqref{eq:gamma2} solution has the form
\begin{align}
&p=-2iK_0\dfrac{H(u_1+i\delta,u_2+1)}{H(u_1,u_2)}
\exp\{2iK_1x+2iK_2t\},  \label{eq:sol}\\
&\abs{p}^2=-4K^2_0\dfrac{H(u_1-i\delta,u_2-1)H(u_1+i\delta,u_2+1)} {H^2(u_1,u_2)}, \label{eq:sol2}
\end{align}
where $u_1=\kappa_1 t+2Z_1$, $u_2=kx+\kappa_2t+2Z_2$,
\begin{gather*}
\kappa_1=4/A_{-},\quad k=2/A_{+},\quad \kappa_2=8\lambda_0/A_{+},\quad \delta=B_{-}^1/A_{-},\\
B_{-}^1=\int_{c^2}^{\infty}\dfrac{dt}{\sqrt{t(t-a^2)(t-b^2)(t-c^2)}},\\
H(u_1,u_2)=\vartheta_3(u_1|2i\frb_{-})\vartheta_3(u_2|2i\frb_{+})+
\vartheta_2(u_1|2i\frb_{-})\vartheta_3(u_2|2i\frb_{+})+\\
+\vartheta_3(u_1|2i\frb_{-})\vartheta_2(u_2|2i\frb_{+})-
\vartheta_2(u_1|2i\frb_{-})\vartheta_2(u_2|2i\frb_{+}).
\end{gather*}
The coefficient $K_0$ equals
\begin{equation*}
K_0=ic\exp(D_{-}\delta-F_{-}),
\end{equation*}
where
\begin{gather*}
D_{-}=\dfrac12\int_0^{a^2}\dfrac{tdt}{\sqrt{t(a^2-t)(b^2-t)(c^2-t)}},\\
F_{-}=\dfrac12\int_{c^2}^\infty\left(\dfrac{t}{\sqrt{t(t-a^2)(t-b^2)(t-c^2)}}-\dfrac1t\right)dt.
\end{gather*}

\section{Investigation of a dependence of solution \eqref{eq:sol} on the parameters of the spectral curve}

If $\lambda_0=0$, then variables $x$ and $t$ are separated to different phases of solution \eqref{eq:sol}. Therefore, for $\lambda_0=0$ an amplitude of solution \eqref{eq:sol} is a periodic function in $x$ and in $t$ with periods $X=A_{+}/2$ and $T=A_{-}/4$. If $\lambda_0\ne0$, then the variable $t$ is situated in two phases with periods $T$ and $T'=A_{+}/(8\lambda_0)=X/(4\lambda_0)$. Thus, if $\lambda_0\ne0$, then an amplitude of solution \eqref{eq:sol} is periodic function only in $x$. Naturally, if $T$ and $T'$ are commensurate, then an amplitude of solution \eqref{eq:sol} is a periodic function with respect to $x$ as well as to $t$.

We have to notice that in any case the crests of two-phase solutions are placed in nodes of some lattice of  periods.
It follows from properties of hyper-elliptic curve $\Gamma$ \eqref{nls:curve} that real vectors $\bU$ and $\bV$ are linearly independent \cite{Dub81e, Bake}. Therefore, if $g = 2$, then any vector from $\bbR^2$ could be represented as a linear combination of  vectors $\bU$ and $\bV$. In particular this is a true for vectors of periods of Riemann two-dimensional theta function, i.e., for $\be_1=(1,0)^t$ and $\be_2=(0,1)^t$:
\begin{equation*}
\Theta(\bu+\be_j)\equiv\Theta(\bu).
\end{equation*}
Therefore, for any hyper-elliptic curve of genus $g = 2$ there exist real numbers $X_j,T_j$ such that following equality hold
\begin{equation*}
X_j\bU+T_j\bV=\be_j,\quad j=1,2.
\end{equation*}
It follows from this equality and from formula \eqref{eq:abs.p} that an amplitude of solution \eqref{eq:sol} is a periodic function on the plane $XOT$
\begin{equation*}
\abs{p}(x+X_j,t+T_j)\equiv\abs{p}(x,t).
\end{equation*}
This proposition is independent from the fact that two-dimensional theta function is expressed by elliptic functions or not. Since a real vector of initial phase $\bZ$ could be decomposed on vectors $\bU$ and  $\bV$, then a change of initial real phase of two-phase solution leads to the trivial shift of a solution on some vector over the plane $XOT$, and it has no influence on the behavior of a solution.  According to this fact in all solutions (except limits) we consider $\bZ=\bzero$.

It follows from properties of elliptic theta functions and from formula \eqref{eq:sol2} that the function $\abs{p}^2(u_1,u_2)$:
\begin{enumerate}
\item is a two-periodic function in $u_j$,
\begin{multline*}
\abs{p}^2(u_1\pm2,u_2)=\abs{p}^2(u_1,u_2\pm2)=\\
=\abs{p}^2(u_1\pm2i\frb_{-},u_2)=\abs{p}^2(u_1,u_2\pm2i\frb_{+})=\abs{p}^2(u_1,u_2);
\end{multline*}
\item satisfies equalities
\begin{align*}
&\abs{p}^2(u_1\pm i\frb_{-},u_2)=\abs{p}^2(u_1\pm1,u_2)\in\bbR,\\
&\abs{p}^2(u_1,u_2\pm i\frb_{+})=\abs{p}^2(u_1,u_2\pm1)\in\bbR.
\end{align*}
\end{enumerate}
Therefore, there exists a real smooth solution of NLS equation with complex initial phase $\bZ$.
Let us remark that the necessary condition for reality of solution \eqref{eq:sol} is following:
\begin{equation*}
2\Im\bZ=\Im BN,\quad\text{where}\quad N\in\bbZ^g, \quad \Re BN\in\bbZ^g.
\end{equation*}

\begin{figure}[hbtp]
\begin{center}
\parbox{0.49\textwidth}{\includegraphics[width=0.49\textwidth]{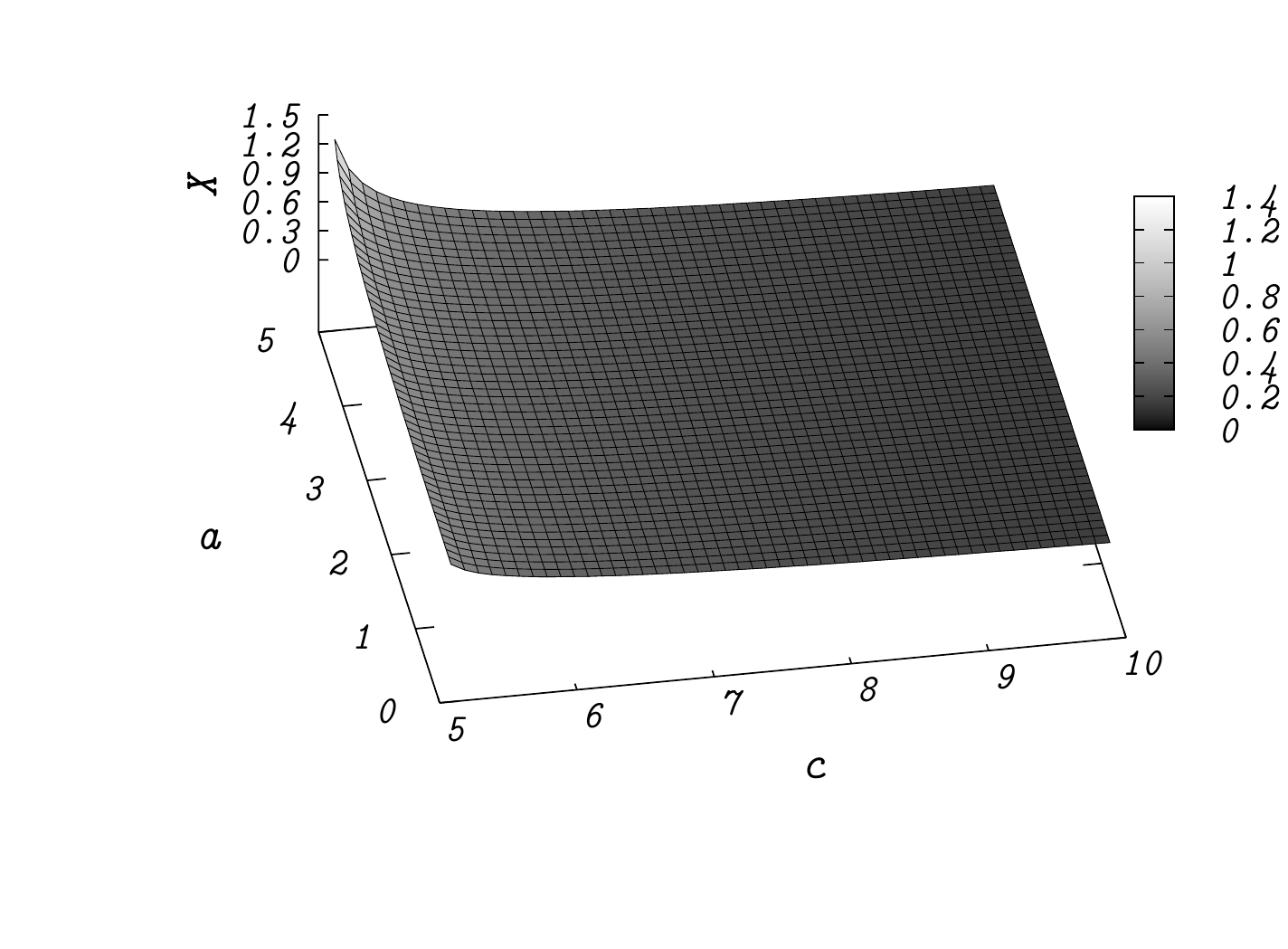}}
\parbox{0.49\textwidth}{\includegraphics[width=0.49\textwidth]{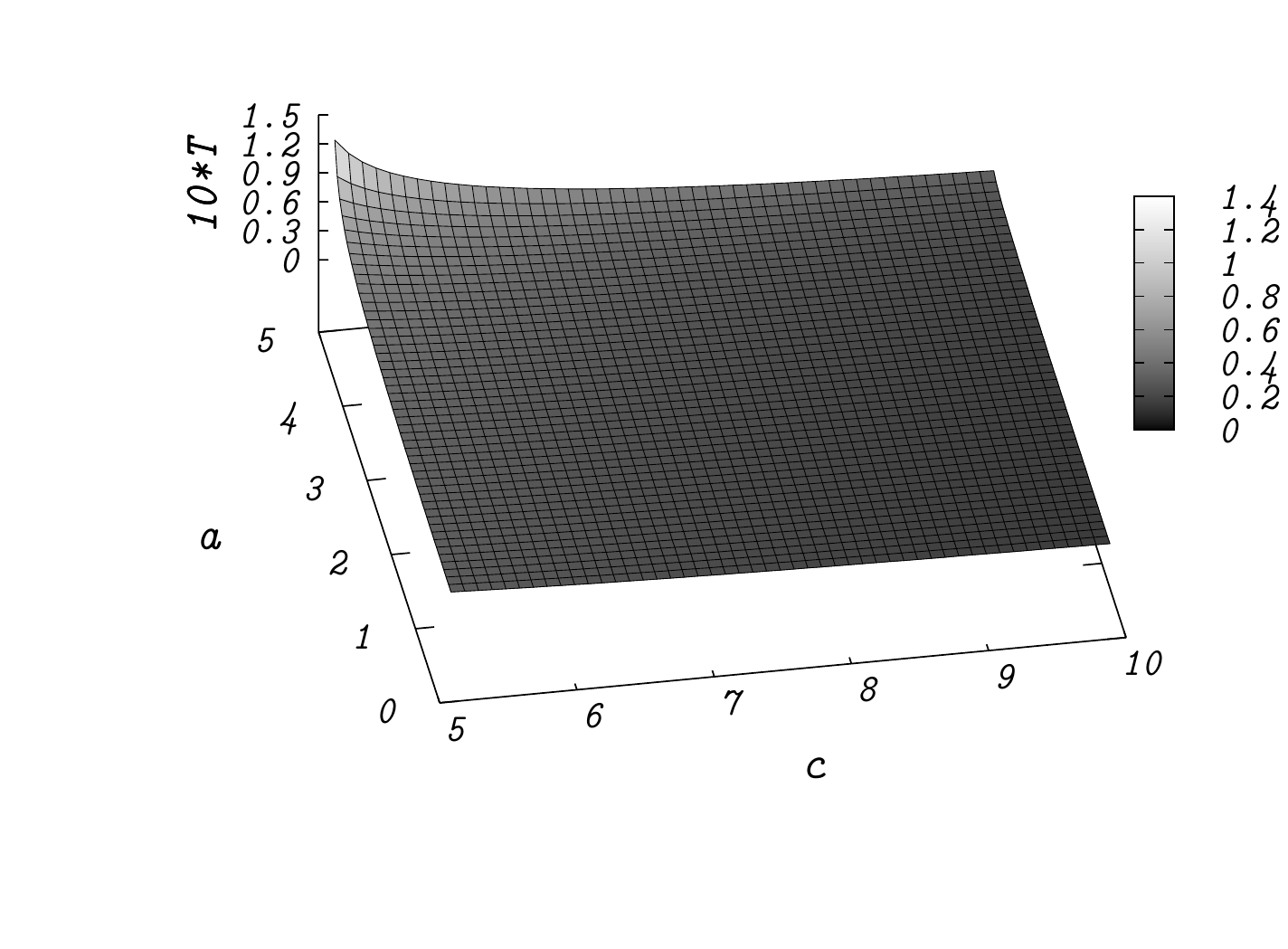}}
\caption{The dependencies of the periods of the phases from $a$ and $c$ when $b=5$}
\label{fig:periods}
\end{center}
\end{figure}

Let us now consider dependencies of the periods $X$ and $T$ from parameters of spectral curve \eqref{eq:gamma2}. We fix the parameter $b$ because a scale transformation of a spectral curve corresponds to scale transformations of solutions' amplitude and periods.
It follows from fig.~\ref{fig:periods} that if a distance between branching points decreases, i.e., if $a\to b$ or $c\to b$, then periods $X$ and $T$ grow. In contrary, if $a$ decreases and $c$ grows, then the periods of the phases of solution \eqref{eq:sol} approach to zero ($X\to0$ and $T\to0$).

\begin{figure}[hbt]
\begin{center}
\parbox{0.49\textwidth}{\includegraphics[width=0.49\textwidth]{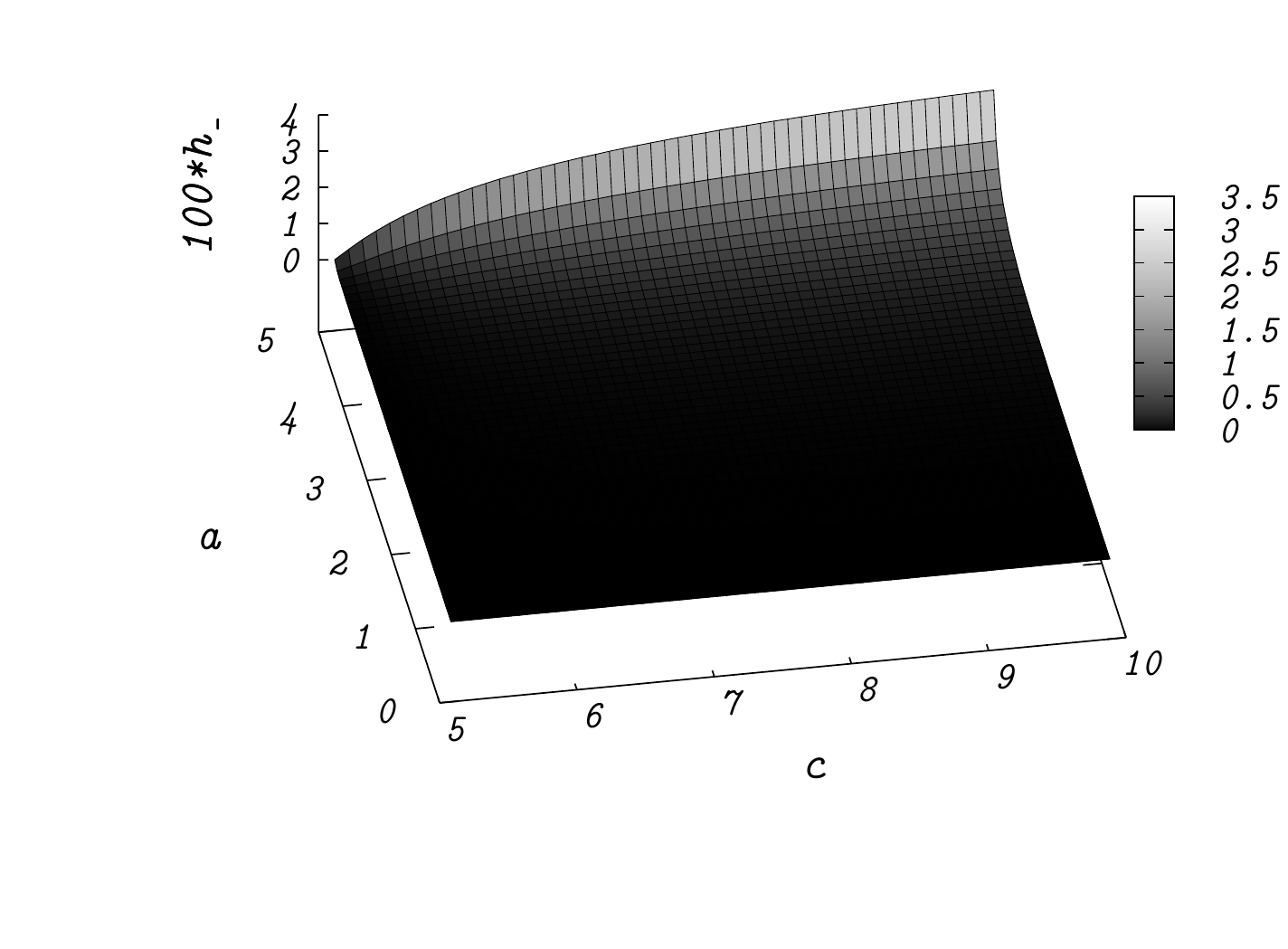}}
\parbox{0.49\textwidth}{\includegraphics[width=0.49\textwidth]{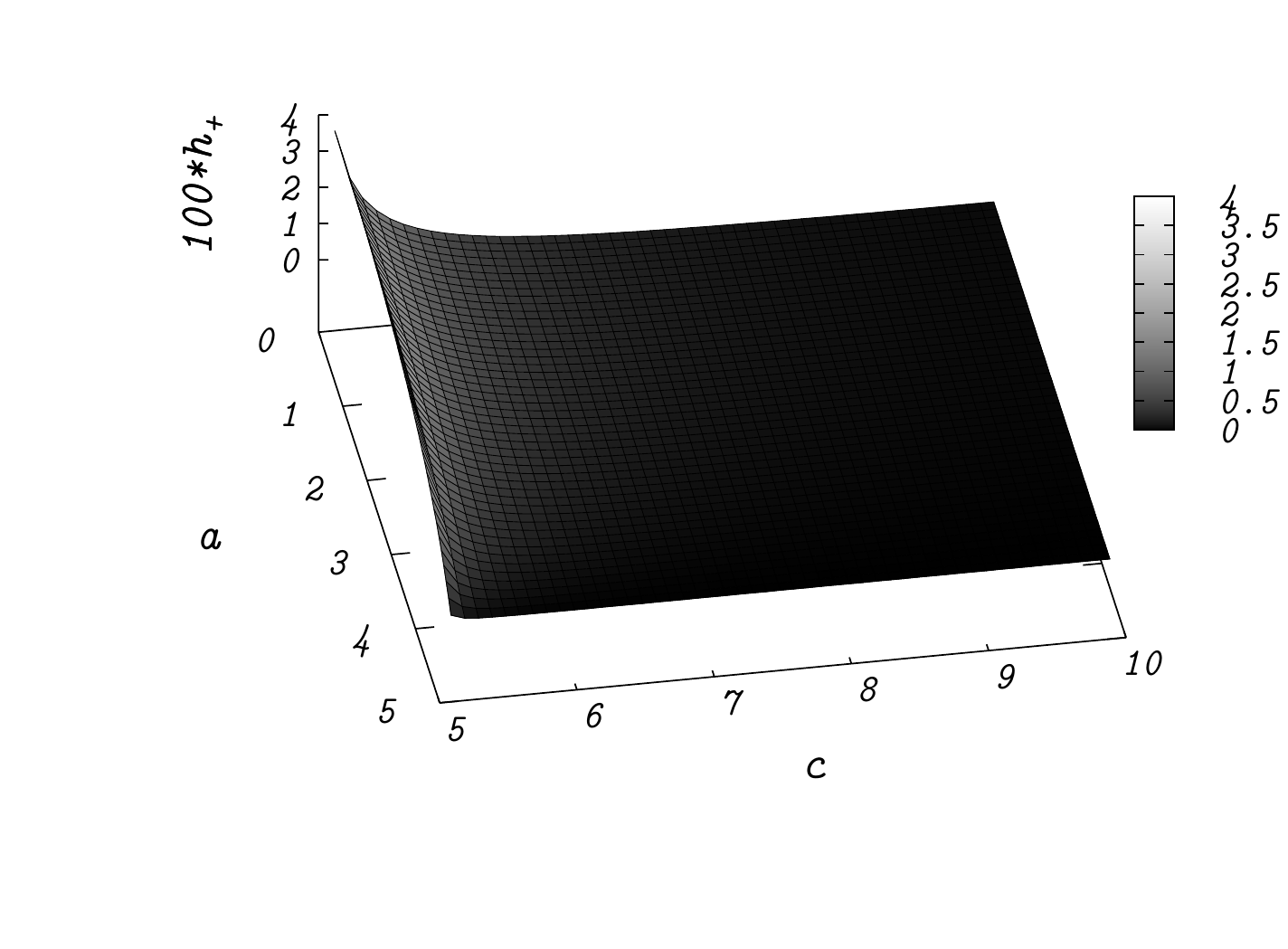}}
\caption{The dependencies of the quantities  $h_{\pm}$ from $a$ and $c$ when $b=5$}
\label{fig:harm}
\end{center}
\end{figure}

It follows from harmonic analyses that a steepness of solution's fronts depends from existence of a harmonics with high numbers in this solution. The larger is a contribution of harmonics with high numbers the larger a steepness of a solution's front. Evidently, main contribution of a highest harmonics in solution is brought by theta functions, though the certain contribution in solution is given by  nonlinearity of the superposition of theta functions. It follows from definition of elliptic Jacobi theta functions that the close to zero are  $h_{\pm}=\exp\{-2\pi\frb_{\pm}\}$  the less a contribution of high harmonics of a corresponding phase in solution.  The dependence of the quantities $h_{\pm}$ from the parameters of spectral curve \eqref{eq:gamma2} is shown on fig.~\ref{fig:harm}.
It follows from fig.~\ref{fig:harm} that for $a/b\to 1$ the first phase ($\kappa_1t$) becomes more expressed, and for $a/b\to0$ it is less noticeable. Also, if $c\to b$ and $a\to b$, then an amplitude of solution is similar to periodically disposed freak waves because a presence of high harmonics in first and second phases.

\begin{figure}[hbtp]
\begin{center}
\parbox{0.49\textwidth}{\includegraphics[width=0.49\textwidth]{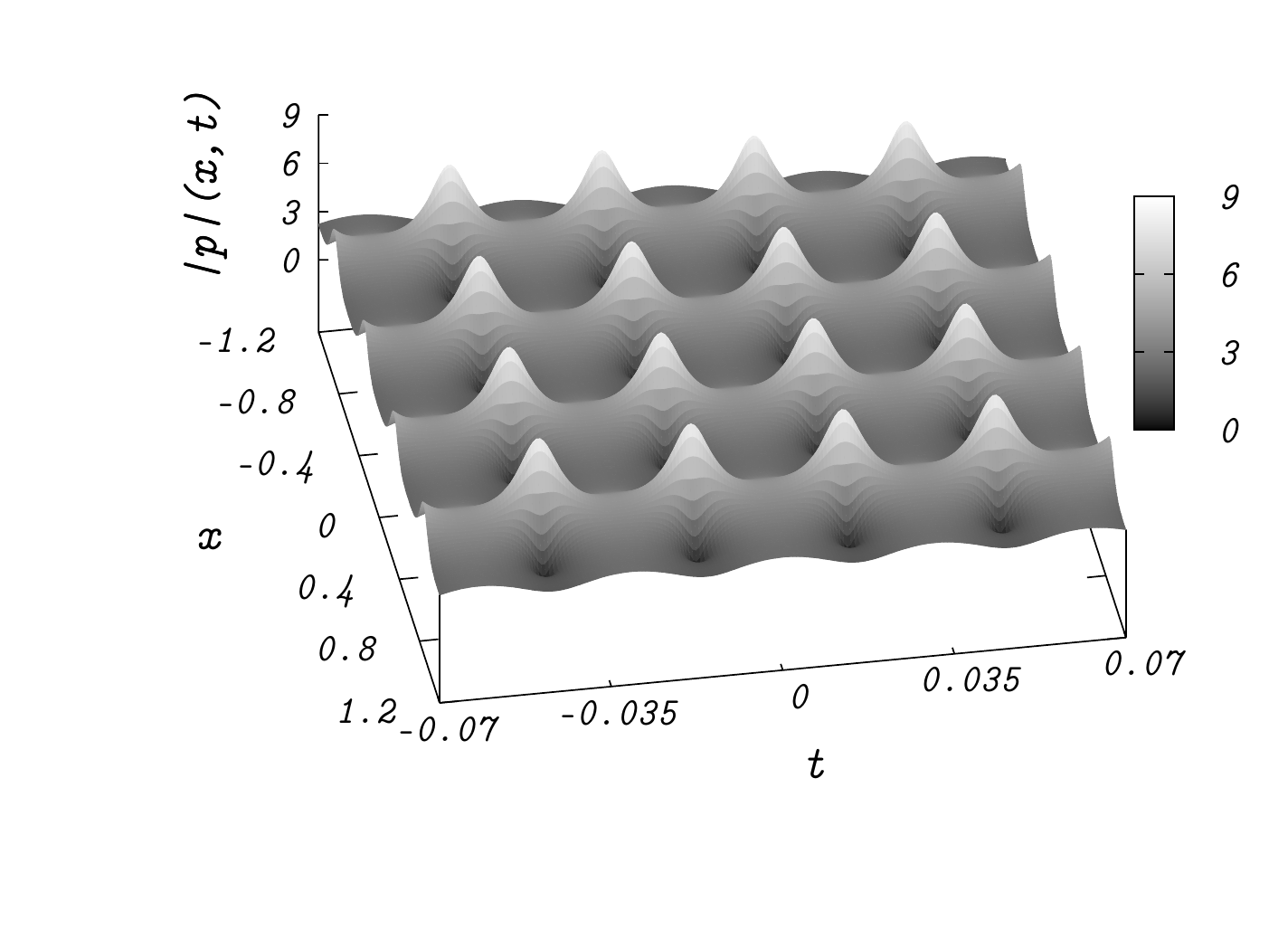}}
\parbox{0.49\textwidth}{\includegraphics[width=0.49\textwidth]{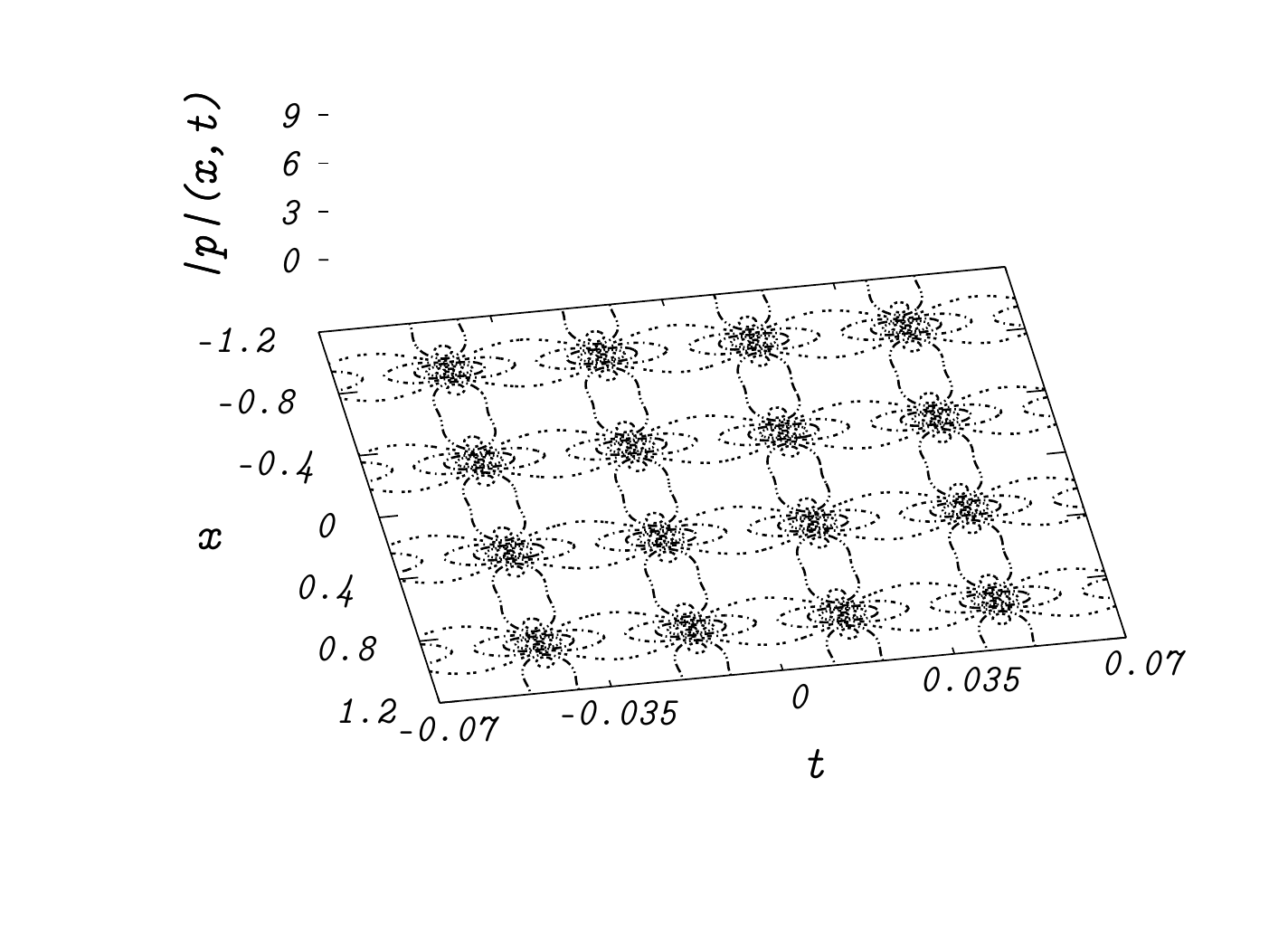}}
\caption{Amplitude of solution \eqref{eq:sol} for $\lambda_0=0$, $a=6$, $b=8$, $c=9$.}
\label{fig:sol.1}
\end{center}
\end{figure}

\begin{figure}[hbt]
\begin{center}
\parbox{0.49\textwidth}{\includegraphics[width=0.49\textwidth]{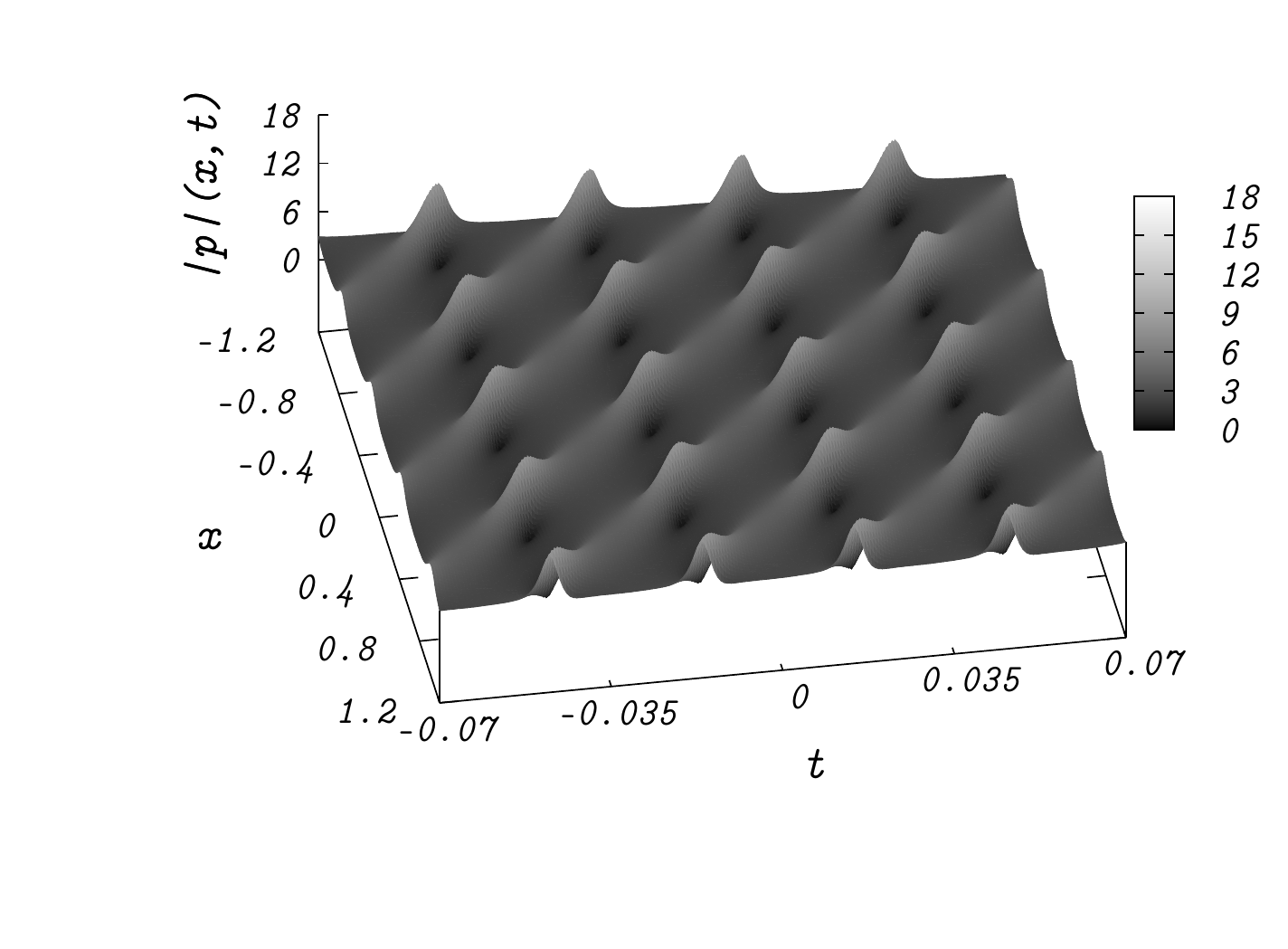}}
\parbox{0.49\textwidth}{\includegraphics[width=0.49\textwidth]{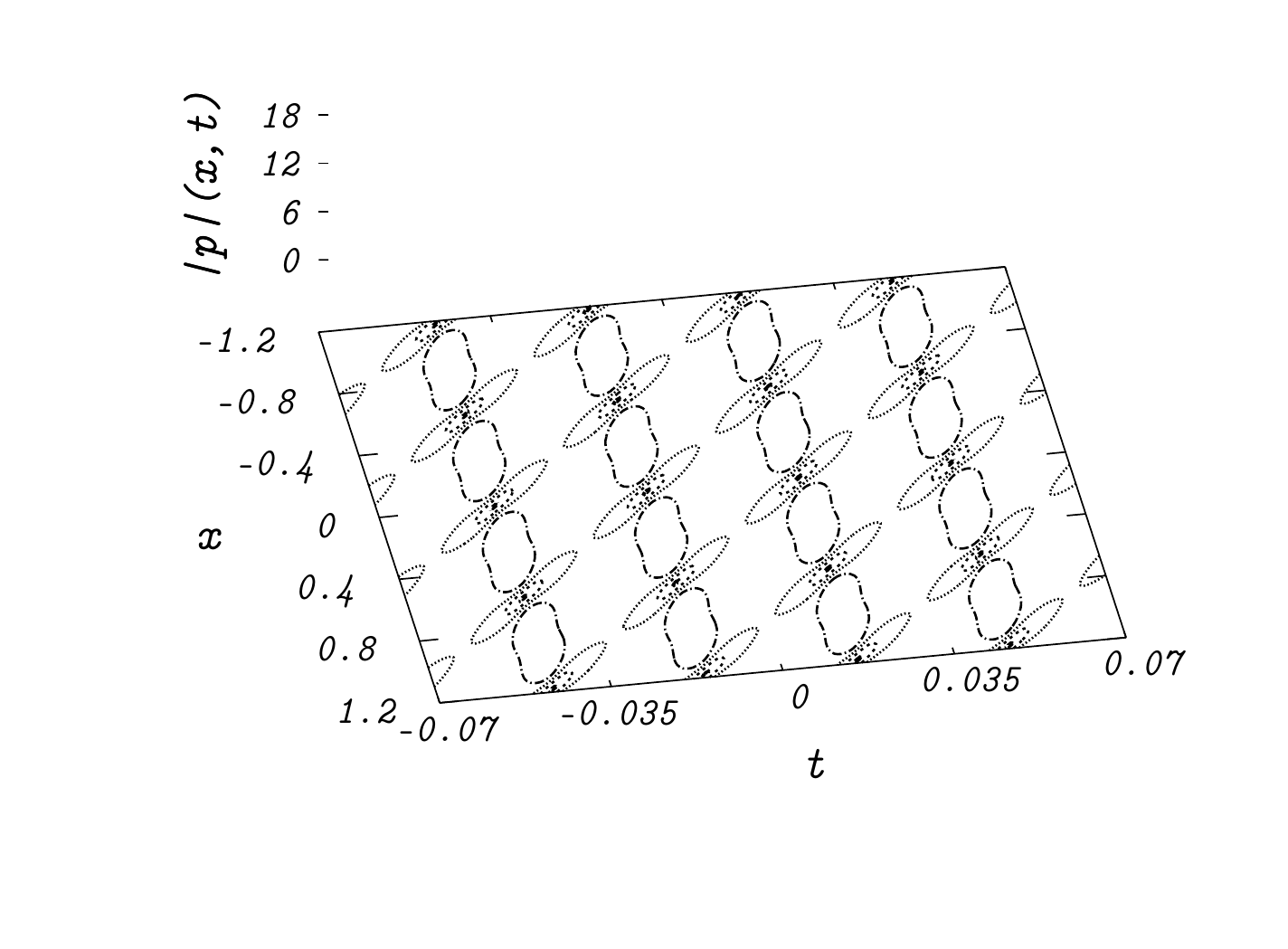}}
\caption{Amplitude of solution \eqref{eq:sol} for $\lambda_0=A_{+}/(2A_{-})$, $a=6$, $b=8$, $c=9$.}
\label{fig:sol.2}
\end{center}
\end{figure}

\begin{figure}[hbt]
\begin{center}
\parbox{0.49\textwidth}{\includegraphics[width=0.49\textwidth]{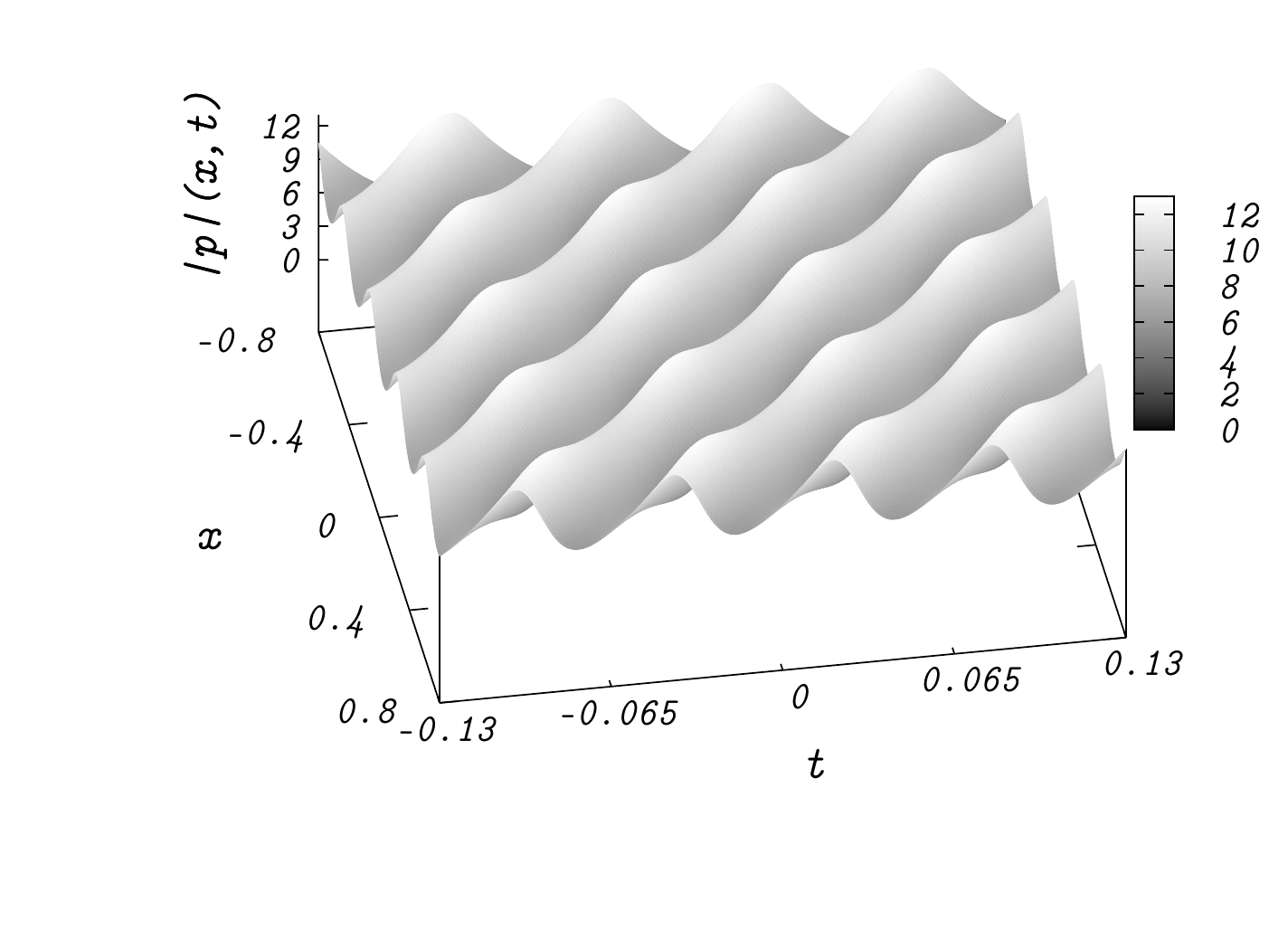}}
\parbox{0.49\textwidth}{\includegraphics[width=0.49\textwidth]{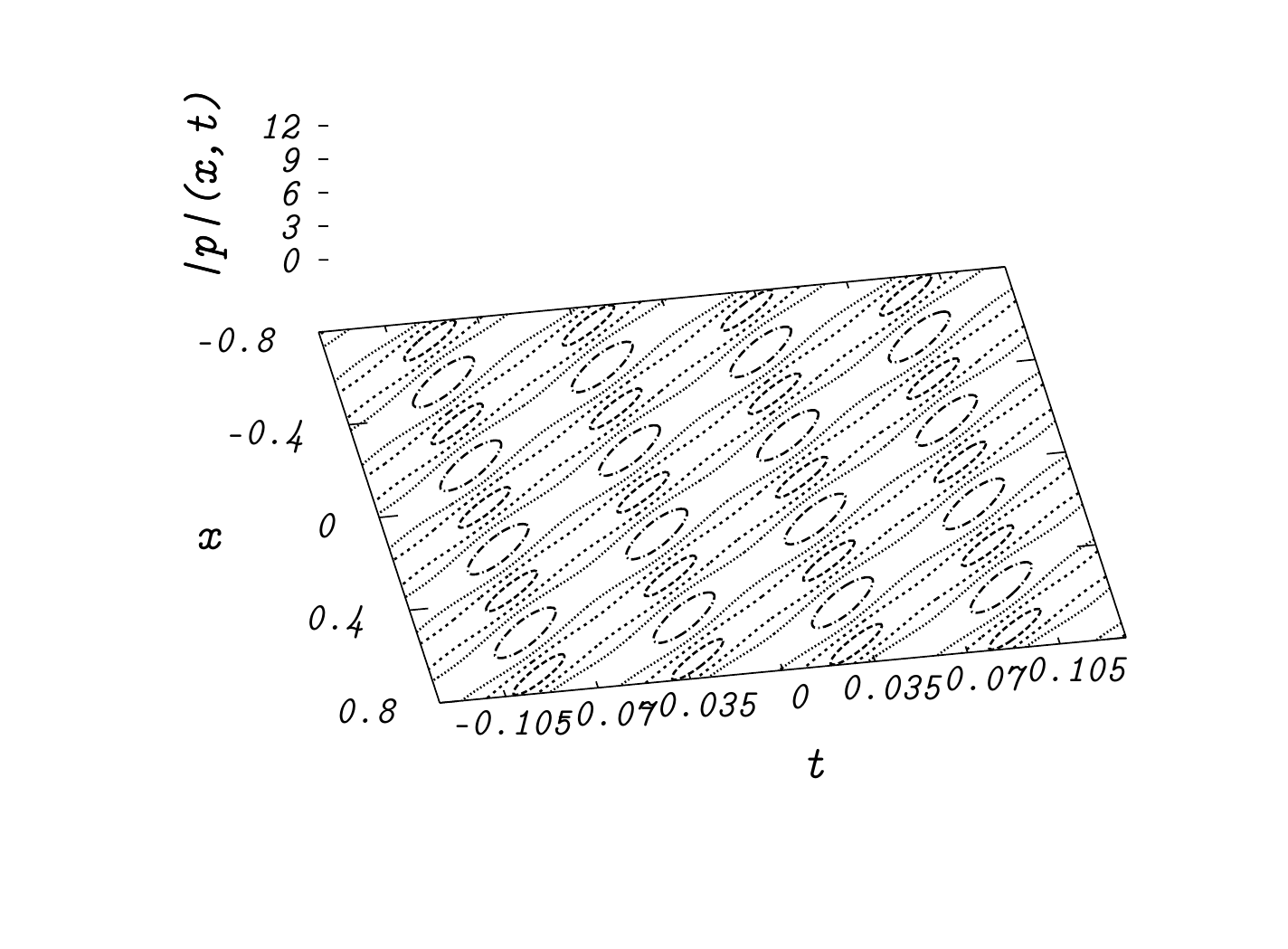}}
\caption{Amplitude of solution \eqref{eq:sol} for $\lambda_0=A_{+}/(2A_{-})$, $a=1$, $b=3$, $c=9$.}
\label{fig:sol.3}
\end{center}
\end{figure}

In the end of this section let us present several figures.  The fig.~\ref{fig:sol.1} gives an example of a periodic in $x$ and in $t$ solution with behavior of periodically disposed ``freak waves''  because of both phases of a solution have steep fronts ($a\approx b$ and $c\approx b$).
The examples of amplitudes of periodic solutions for $T'=T$ and for different values of parameters of a spectral curve are presented on figures ~\ref{fig:sol.2} and~\ref{fig:sol.3}.  It is easy to see how a behavior of two-phase solution \eqref{eq:sol} changes from the case $a\approx b$ and $c\approx b$ (steep fronts, fig.~\ref{fig:sol.2}) to the case $a\ll b$ and $c\gg b$  (sloping fronts, fig.~\ref{fig:sol.3}).

\section{Simplest limits of solutions}

In this section we consider several limits of solution \eqref{eq:sol}; these limits correspond to different confluences of branch points of spectral curve \eqref{eq:gamma2}.

Let $c\to b$ in solutions \eqref{eq:sol}. Then a limit of solution \eqref{eq:sol} has the form of the plane wave:
\begin{multline}
p(x,t)\Big|_{c=b}=2\dfrac{(1+\sqrt{1-k_a^2})^2}{2k_a}\cdot
\left(\dfrac{1 +\sqrt{1-k_a^2}}{k_a}\right)^{-2}\exp(-i\kappa_{1}'t)\times\\
\times \exp\{-2i\lambda_0 x+2i(-2\lambda_0^2+a^2+i\kappa'_1/2)t\}=\\
=a \exp\{-2i\lambda_0 x-2i(2\lambda_0^2-a^2)t\}. \label{eq:sol.deg1}
\end{multline}

If $a\to b$, then a limit of solution takes the form of the another plane wave
\begin{equation}
p(x,t)\Big|_{a=b}=c\exp\{-2i\lambda_0 x-2i(2\lambda_0^2-c^2)t-i\vphi/2\}. \label{eq:sol.deg3}
\end{equation}

\begin{figure}[hbtp]
\begin{center}
\parbox{0.49\textwidth}{\includegraphics[width=0.49\textwidth]{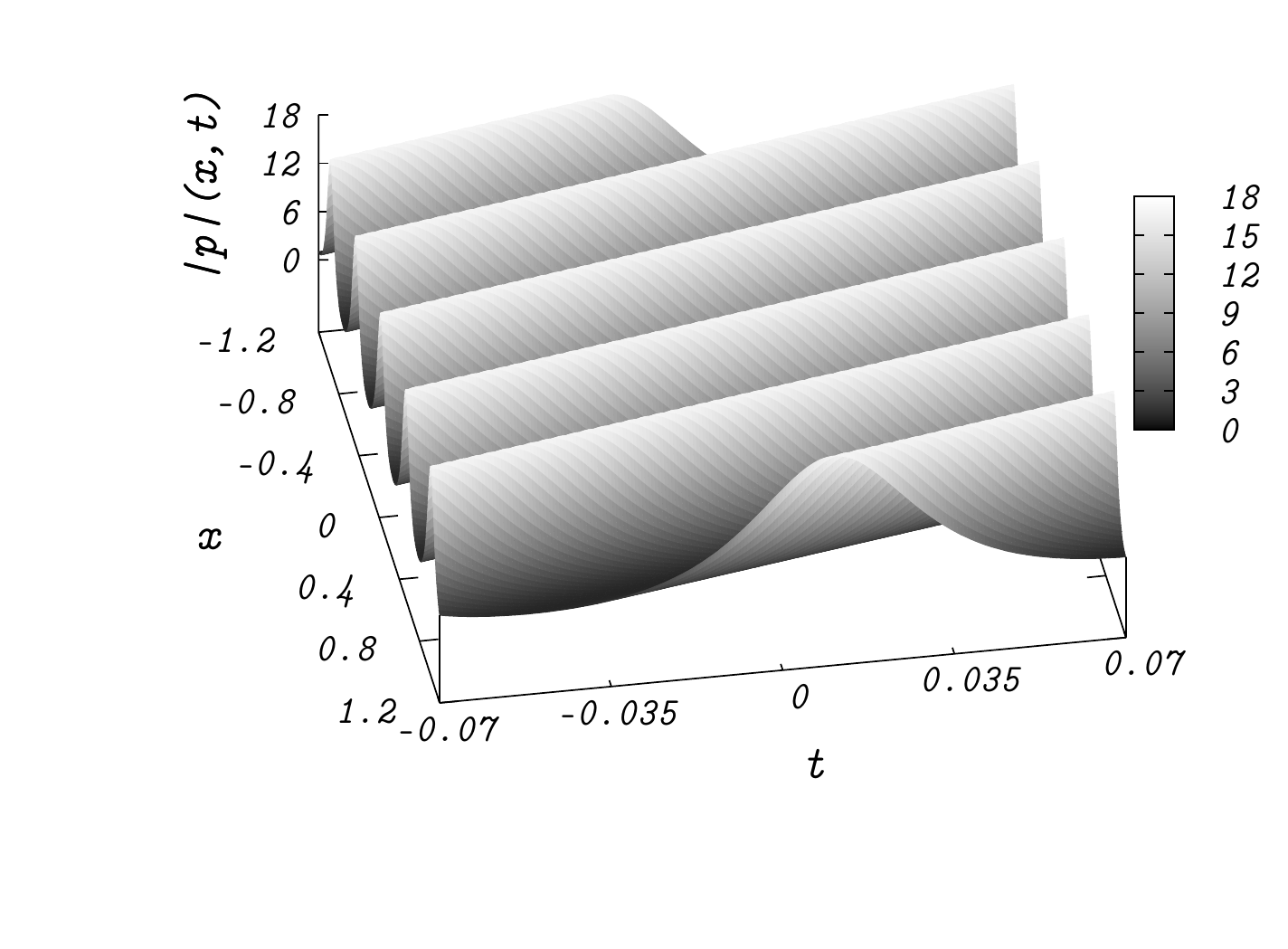}}
\parbox{0.49\textwidth}{\includegraphics[width=0.49\textwidth]{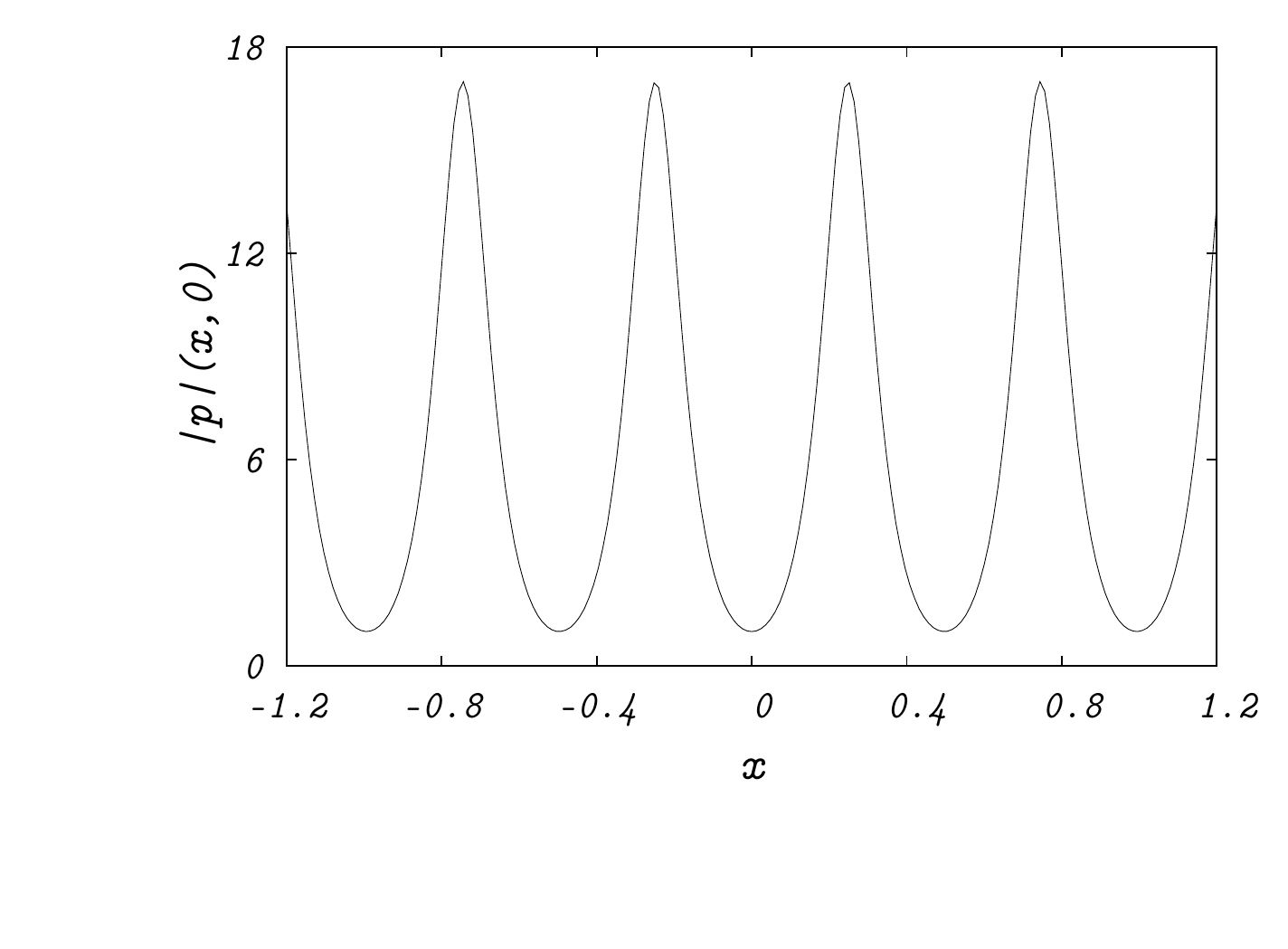}}
\caption{The amplitude of the degenerated solution  \eqref{eq:sol.deg2} when $\lambda_0=1$, $b=8$, $c=9$.}
\label{fig:sol.deg2}
\end{center}
\end{figure}

On the other hand, a limit of solution \eqref{eq:sol} as $a\to 0$ equals one-phase traveling wave
(fig.~\ref{fig:sol.deg2})
\begin{multline}
p(x,t)\Big|_{a=0}=\sqrt{c^2-b^2}\dfrac{\vartheta_3(kx+\kappa_2t|2ib_{+})-\vartheta_2(kx+\kappa_2t|2ib_{+})} {\vartheta_3(kx+\kappa_2t|2ib_{+})+\vartheta_2(kx+\kappa_2t|2ib_{+})}\times\\
\times\exp\{2i(K_1x+K_2t)\}.
\label{eq:sol.deg2}
\end{multline}

In order to obtain another representation of solution \eqref{eq:sol.deg2} we rewrite a solution in the form
\begin{equation}
p(x,t)\Big|_{a=0}=f(x+4\lambda_0t)e^{-2i\lambda_0x+2i(K_{20}-2\lambda_0^2)t},\quad f(x)\in\bbR,\quad K_{20}=b^2+c^2. \label{eq:sol.deg2.2}
\end{equation}
Placing this expression into the equation \eqref{eq:nls} and simplifying it, obtain a differential equation with respect to $f(x)$
\begin{equation*}
-2K_{20}f+f_{xx}+2f^3=0
\end{equation*}
or
\begin{equation*}
f_x^2=C+2K_{20}f^2-f^4.
\end{equation*}
A solution of last equation is an elliptic Jacobi function \cite{Akhe}.
It follows from the fig.~\ref{fig:sol.deg2} that a function $f(x)$ from \eqref{eq:sol.deg2.2} equals
\begin{equation*}
f(x)=A\dn(B(x-x_0);\wkk),
\end{equation*}
where parameters $A$, $B$ and $\wkk$ satisfy the relations
\begin{equation*}
A=B=\sqrt{\dfrac{2-\wkk^2}{2K_{20}}},\quad C=\wkk^2-1.
\end{equation*}
Therefore, the solution \eqref{eq:sol.deg2} may be rewritten in the form (see also \cite{IAKe, AK85e})
\begin{equation*}
p(x,t)=A\dn(A(x+4\lambda_0t-x_0);\wkk)e^{-2i\lambda_0x+2i(K_{20}-2\lambda_0^2)t}.
\end{equation*}

It is not difficult to see that limits of solutions correspond to solutions that is constructed on limits of spectral curves. The solution \eqref{eq:sol.deg1} corresponds to the rational spectral curve
\begin{equation*}
w^2=(\lambda^2-2\lambda_0\lambda+|\lambda_1|^2),
\end{equation*}
and \eqref{eq:sol.deg3} corresponds to the curve (also of genus $g=0$)
\begin{equation*}
w^2=(\lambda^2-2\lambda_0\lambda+|\lambda_3|^2).
\end{equation*}
The one-phase solution \eqref{eq:sol.deg2} is associated with the curve of genus $g=1$
\begin{equation*}
w^2=(\lambda^2-2\lambda_0\lambda+|\lambda_2|^2)(\lambda^2-2\lambda_0\lambda+|\lambda_3|^2).
\end{equation*}

Some technical details of related calculations can be found in Appendix.

%

\section*{Concluding remarks}

From the present work and from previous papers \cite{Sm12tmfe, Sm13mze} it follows that:
a) if an involution $\tau:(w,\lambda)\to(w,-\lambda)$ exists on a spectral curve $\Gamma=\{(w,\lambda)\}$, then two-gap solution of NLS equation is a periodic in $x$ and in $t$ function;
b) if a distance between branching points decreases, then a steepness of front of solution increases.
Let us remark that two-phase periodic in $x$ and in $t$ solutions from the present work and from \cite{Sm13mze} are define by different formulae; also these solutions have different shapes.
The analysis of simplest limits of smooth multi-phase solutions \eqref{eq:p,q} of NLS equation show us that simplest confluences of branching points lead to solutions in the form of plane waves or traveling waves. In order to obtain more interesting solutions it is necessary to consider scaling limits with confluence several pairs of branching points.

Authors acknowledge Prof. V.B. Matveev for his support and discussion of this paper. Prof. V.Zinger wishes to express his gratitude the support and hospitality of the St. Petersburg State University of Aerospace Instrumentation where he held the visiting position in the summer of 2013, which has made this collaboration possible. This work performed within the framework of the state order of the Ministry of education of Russia, and partially supported by RFBR, research project ¹14-01-00589\_a.

\appendix

\section{Asymptotic of the parameters of the solutions}

\subsection{Limit of \eqref{eq:sol} as  $c\to b$}

Let $a=k_ab$, $0<k_a<1$, $c=b+\epsilon$, $\epsilon\to+0$. Then
\begin{align*}
&A_{+}=\dfrac{2K\left(\sqrt{\dfrac{b^2-a^2}{c^2-a^2}}\right)}{\sqrt{c^2-a^2}}
\sim\dfrac{-1}{b\sqrt{1-k_a^2}}\ln\dfrac{\epsilon}{8b(1-k_a^2)}\to+\infty,\\
&B_{+}=\dfrac{2K\left(\sqrt{\dfrac{c^2-b^2}{c^2-a^2}}\right)}{\sqrt{c^2-a^2}}
\sim\dfrac{\pi}{b\sqrt{1-k_a^2}},\\
&A_{-}=\dfrac{2K\left(\dfrac{a}{b}\sqrt{\dfrac{c^2-b^2}{c^2-a^2}}\right)}{b\sqrt{c^2-a^2}}
\sim\dfrac{\pi}{b^2\sqrt{1-k_a^2}},\\
&B_{-}=\dfrac{2K\left(\dfrac{c}{b}\sqrt{\dfrac{b^2-a^2}{c^2-a^2}}\right)}{b\sqrt{c^2-a^2}}
\sim\dfrac{-1}{b^2\sqrt{1-k_a^2}}\ln\dfrac{k_a^2\epsilon}{8b(1-k_a^2)}\to+\infty,\\
&B_{-}^1=\dfrac{2F\left(\dfrac{b}{c},\dfrac{c}{b}\sqrt{\dfrac{b^2-a^2}{c^2-a^2}}\right)}{b\sqrt{c^2-a^2}}=\\
&=\dfrac{2\left[K\left(\dfrac{c}{b}\sqrt{\dfrac{b^2-a^2}{c^2-a^2}}\right)
-F\left(\dfrac{\sqrt{c^2-a^2}}{c},\dfrac{c}{b}\sqrt{\dfrac{b^2-a^2}{c^2-a^2}}\right)\right]}{b\sqrt{c^2-a^2}}
\sim\\
&\sim\dfrac{-1}{b^2\sqrt{1-k_a^2}}\ln\dfrac{k_a(1+\sqrt{1-k_a^2})\epsilon}{8b(1-k_a^2)}\to+\infty,\\
&D_{-}=\dfrac{c^2\left[K\left(\dfrac{a}{b}\sqrt{\dfrac{c^2-b^2}{c^2-a^2}}\right)
-\Pi\left(\dfrac{a^2}{a^2-c^2},\dfrac{a}{b}\sqrt{\dfrac{c^2-b^2}{c^2-a^2}}\right)\right]}{b\sqrt{c^2-a^2}}
\sim\dfrac{\pi(1-\sqrt{1-k_a^2})}{2\sqrt{1-k_a^2}},\\
&F_{-}\sim \ln\dfrac{2}{1+\sqrt{1-k_a^2}}+\dfrac{b^2}2B_{-}^1,
\end{align*}
where $K(m)$, $F(m)$ and $\Pi(m)$ are canonical elliptic integrals.

Therefore, the parameters of the solution \eqref{eq:sol} have following asymptotic as $c\to b$:
\begin{gather*}
\frb_{-}=\dfrac{B_{-}}{A_{-}}\to+\infty,\quad
\kappa_1\sim\dfrac{4b^2\sqrt{1-k_a^2}}{\pi},\quad
\delta\sim\frb_{-}-\dfrac2{\pi}\ln\dfrac{1+\sqrt{1-k_a^2}}{k_a},\\
\frb_{+}=\dfrac{B_{+}}{A_{+}}\to0,\quad k=\dfrac2{A_{+}}\to0,\quad \kappa_2=\dfrac{8\lambda_0}{A_{+}}\to0,\\
K_0\sim i\dfrac{(1+\sqrt{1-k_a^2})^2}{2k_a}\exp(-\pi\frb_{-}/2)\to0.
\end{gather*}

Passing to theta functions that define the first phase of solution \eqref{eq:sol} we obtain
\begin{align*}
&\vartheta_3(\kappa_1t|2i\frb_{-})\sim1,\quad
\vartheta_2(\kappa_1t|2i\frb_{-})\sim2h_{-}\cos(\kappa_{1}'t),\\
&\vartheta_3(\kappa_1t+i\delta|2i\frb_{-})\sim1
+\left(\dfrac{1+\sqrt{1-k_a^2}}{k_a}\right)^{-4}e^{-2i\kappa_{1}'t},\\
&\vartheta_2(\kappa_1t+i\delta|2i\frb_{-})\sim h_{-}^{-1}
\left(\dfrac{1+\sqrt{1-k_a^2}}{k_a}\right)^{-2}e^{-i\kappa_{1}'t},\\
&\vartheta_3(\kappa_1t-i\delta|2i\frb_{-})\sim1
+\left(\dfrac{1+\sqrt{1-k_a^2}}{k_a}\right)^{-4}e^{2i\kappa_{1}'t},\\
&\vartheta_2(\kappa_1t-i\delta|2i\frb_{-})\sim h_{-}^{-1}
\left(\dfrac{1+\sqrt{1-k_a^2}}{k_a}\right)^{-2}e^{i\kappa_{1}'t},
\end{align*}
where $ h_{-}=\exp(-\pi\frb_{-}/2)$, $\kappa_1'=4b^2\sqrt{1-k_a^2}$.

To obtain an asymptotic of second phase we use following relations  \cite{Akhe}:
\begin{equation}
\begin{gathered}
\vartheta_2(u|i\frb)=\dfrac1{\sqrt{\frb}}e^{-\pi u^2/\frb}\vartheta_4\left(\dfrac{iu}\frb\right.\left|\dfrac{i}\frb\right),\\
\vartheta_3(u|i\frb)=\dfrac1{\sqrt{\frb}}e^{-\pi
u^2/\frb}\vartheta_3\left(\dfrac{iu}\frb\right.\left|\dfrac{i}\frb\right)
\end{gathered} \label{eq:app.1}
\end{equation}

Let us set a nonzero initial phase  $\bZ=(0,1/4)^t$.
As the result we get
\begin{align*}
&\vartheta_3(kx+\kappa_2t+2Z_2\pm1|2i\frb_{+})=\vartheta_3(kx+\kappa_2t+2Z_2|2i\frb_{+}),\\
&\vartheta_2(kx+\kappa_2t+2Z_2\pm1|2i\frb_{+})=-\vartheta_2(kx+\kappa_2t+2Z_2|2i\frb_{+}),\\
&\vartheta_3(kx+\kappa_2t+2Z_2|2i\frb_{+})\sim h_{+}\left(1+e^{k'(x+4\lambda_0t)}\right),\\
&\vartheta_2(kx+\kappa_2t+2Z_2|2i\frb_{+})\sim h_{+}\left(1-e^{k'(x+4\lambda_0t)}\right),
\end{align*}
where
\begin{equation*}
h_{+}=\dfrac{1}{\sqrt{2\frb_{+}}}\exp\left\{-\dfrac{\pi}{2\frb_{+}}(kx+\kappa_2t+2Z_2)^2\right\},\quad
k'=2b\sqrt{1-k_a^2}.
\end{equation*}

It can easily be checked that
\begin{equation*}
K_1\sim-\lambda_0,\quad
K_2\sim-2\lambda_0^2+a^2+2b^2\sqrt{1-k_a^2} .
\end{equation*}

\subsection{Limit of \eqref{eq:sol} as  $a\to b$}

If $a\to b$, then asymptotic of integrals describes by next relations:
\begin{align*}
&A_{+}\sim\dfrac1{\sqrt{c^2-b^2}}\int_{a^2}^{b^2}\dfrac{dt}{\sqrt{(t-a^2)(b^2-t)}} =\dfrac{\pi}{\sqrt{c^2-b^2}},\\
&B_{+}\sim\int_{b^2}^{c^2}\dfrac{dt}{(t-b^2)\sqrt{(c^2-t)}}\to+\infty,\\
&A_{-}\sim\int_0^{b^2}\dfrac{dt}{(b^2-t)\sqrt{t(c^2-t)}}\to+\infty,\\
&B_{-}\sim\dfrac1{b\sqrt{c^2-b^2}}\int_{a^2}^{b^2}\dfrac{dt}{\sqrt{(t-a^2)(b^2-t)}} =\dfrac{\pi}{b\sqrt{c^2-b^2}},\\
&B_{-}^1\sim\int_{c^2}^{\infty}\dfrac{dt}{(t-b^2)\sqrt{t(t-c^2)}}
=\dfrac1{b\sqrt{c^2-b^2}}\arccos\dfrac{c^2-2b^2}{c^2},\\
&D_{-}\sim\dfrac12\int_{0}^{b^2}\dfrac{tdt}{(b^2-t)\sqrt{t(c^2-t)}}=
\dfrac{b^2}2A_{-}-\dfrac12\arccos\dfrac{c^2-2b^2}{c^2},\\
&F_{-}\sim\dfrac12\int_{c^2}^{\infty}\left(\dfrac1{\sqrt{t(t-c^2)}}-\dfrac1t\right)dt+
\dfrac{b^2}2\int_{c^2}^{\infty}\dfrac{dt}{(t-b^2)\sqrt{t(t-c^2)}}=\\
&\quad\quad= \ln2+\dfrac{b^2}2B_{-}^1.
\end{align*}

Therefore, in this case we have:
\begin{gather*}
\frb_{-}=\dfrac{B_{-}}{A_{-}}\to0,\quad
\kappa_1=\dfrac4{A_{-}}\to0,\quad
\delta=\dfrac{B_{-}^1}{A_{-}}\to0,\quad \frb_{+}=\dfrac{B_{+}}{A_{+}}\to+\infty,\\
k\sim \dfrac{2\sqrt{c^2-b^2}}{\pi},\quad
\kappa_2\sim\dfrac{8\lambda_0\sqrt{c^2-b^2}}{\pi},\quad
K_0\sim i\dfrac{c}2.
\end{gather*}

Asymptotic of theta functions that define first phase of solution \eqref{eq:sol} describes by following relations
\begin{align*}
&\vartheta_3(\kappa_1t+2Z_1|2i\frb_{-})\sim\wth_{-}(t)\left(1+e^{\wtk_1t}\right),\\
&\vartheta_2(\kappa_1t+2Z_1|2i\frb_{-})\sim\wth_{-}(t)\left(1-e^{\wtk_1t}\right),\\
&\vartheta_3(\kappa_1t+2Z_1+i\delta|2i\frb_{-})\sim\wth_{-}(t+iB_{-}^1/4)\left(1+e^{\wtk_1t+i\vphi}\right),\\
&\vartheta_2(\kappa_1t+2Z_1+i\delta|2i\frb_{-})\sim\wth_{-}(t+iB_{-}^1/4)\left(1-e^{\wtk_1t+i\vphi}\right),\\
&\vartheta_3(\kappa_1t+2Z_1-i\delta|2i\frb_{-})\sim\wth_{-}(t-iB_{-}^1/4)\left(1+e^{\wtk_1t-i\vphi}\right),\\
&\vartheta_2(\kappa_1t+2Z_1-i\delta|2i\frb_{-})\sim\wth_{-}(t-iB_{-}^1/4)\left(1-e^{\wtk_1t-i\vphi}\right),
\end{align*}
where
$\wtk_1=4b\sqrt{c^2-b^2}$, initial phase $\bZ=(1/4,0)^t$,
\begin{equation*}
\wth_{-}(t)=\dfrac{1}{\sqrt{2\frb{-}}}\exp\left\{-\dfrac{\pi}{2\frb_{-}}(\kappa_1t+2Z_1)^2\right\},\quad
\vphi=\arccos\dfrac{c^2-2b^2}{c^2}.
\end{equation*}
By calculating these asymptotic we use again formulas \eqref{eq:app.1}.

Theta-functions that define second phase of solution \eqref{eq:sol} have the following asymptotic
\begin{align*}
&\vartheta_3(kx+\kappa_2t\pm1|2i\frb_{+})=\vartheta_3(kx+\kappa_2t|2i\frb_{+}),\\
&\vartheta_2(kx+\kappa_2t\pm1|2i\frb_{+})=-\vartheta_2(kx+\kappa_2t|2i\frb_{+}),\\
&\vartheta_3(kx+\kappa_2t|2i\frb_{+})\sim 1,\\
&\vartheta_2(kx+\kappa_2t|2i\frb_{+})\sim 2\wth_{+}\cos(\widetilde{k}x+\wtk_{2}t),
\end{align*}
where $\wth_{+}=\exp(-\pi\frb_{+}/4)$, $\widetilde{k}=2\sqrt{c^2-b^2}$, $\wtk_{2}=8\lambda_0\sqrt{c^2-b^2}$.
In addition
\begin{equation*}
K_1\sim-\lambda_0,\quad
K_2\sim-2\lambda_0^2+c^2.
\end{equation*}

\subsection{Limit of \eqref{eq:sol} as  $a\to 0$}

If $a\to 0$, then elliptic integrals have following asymptotic:
\begin{align*}
&A_{+}\sim\int_0^{b^2}\dfrac{dt}{\sqrt{t(b^2-t)(c^2-t)}},\quad
B_{+}\sim\int_{b^2}^{c^2}\dfrac{dt}{\sqrt{t(t-b^2)(c^2-t)}},\\
&A_{-}\sim\dfrac{\pi}{bc},\quad
B_{-}\sim\int_0^{b^2}\dfrac{dt}{t\sqrt{(b^2-t)(c^2-t)}}\to+\infty,\\
&B_{-}^1\sim\dfrac1{bc}\ln\dfrac{c+b}{c-b}, \quad
D_{-}\sim0,\quad
F_{-}\sim\dfrac12\ln\dfrac{4c^2}{c^2-b^2}.
\end{align*}

From this asymptotic it follows that calculated on curve $\Gamma_{-}$ parameters have a form:
\begin{equation*}
\frb_{-}=\dfrac{B_{-}}{A_{-}}\to+\infty,\quad
\kappa_1\sim\dfrac{4bc}{\pi},\quad
\delta\sim\dfrac1{\pi}\ln\dfrac{c+b}{c-b},\quad K_0\sim i\dfrac{\sqrt{c^2-b^2}}{2}.
\end{equation*}
Calculated on $\Gamma_{-}$ theta functions have following asymptotic
\begin{align*}
&\vartheta_3(\kappa_1t|2i\frb_{-})\sim1,\\
&\vartheta_2(\kappa_1t|2i\frb_{-})\sim2h_{-}\cos(4bct),\\
&\vartheta_3(\kappa_1t+i\delta|2i\frb_{-})\sim1,\\
&\vartheta_2(\kappa_1t+i\delta|2i\frb_{-})\sim h_{-}\dfrac{(c^2+b^2)\cos(4bct)-2ibc\sin(4bct)}{c^2-b^2} ,\\
&\vartheta_3(\kappa_1t-i\delta|2i\frb_{-})\sim1,\\
&\vartheta_2(\kappa_1t-i\delta|2i\frb_{-})\sim h_{-}\dfrac{(c^2+b^2)\cos(4bct)+2ibc\sin(4bct)}{c^2-b^2}  .
\end{align*}

Calculated on curve $\Gamma_{+}$ parameters have standard form with substitution $a=0$; calculated on $\Gamma_{-}$ theta functions are standard elliptic theta functions.

In addition
\begin{equation*}
K_1\sim-\lambda_0,\quad
K_2\sim-2\lambda_0^2+b^2+c^2.
\end{equation*}


\begin{thebibliography}{10}

\bibitem{AAT}
N.~Akhmediev, A.~Ankiewicz, and M.~Taki.
\newblock Waves that appear from nowhere and disappear without a trace.
\newblock {\em Phys. Lett. A}, 373:675--678, 2009.

\bibitem{EPJ}
N.~Akhmediev and E.~Pelinovsky, editors.
\newblock {\em Discussion \& Debate: Rogue Waves - Towards a Unifying
  Concept?}, volume 185 of {\em Eur. Phys. J. Special Topics}.
\newblock Springer, 2010.

\bibitem{Per83}
D.~H. Peregrine.
\newblock Water waves, nonlinear {S}chr{\"o}dinger equations and their
  solutions.
\newblock {\em J. Austral. Math. Soc. Ser. B}, 25:16--43, 1983.

\bibitem{Zakh08}
A.~I. Dyachenko and V.~E. Zakharov.
\newblock On the formation of freak waves on the surface of deep water.
\newblock {\em JETP Letters}, 88(5):356--359, 2008.

\bibitem{Chab11}
A.~Chabchoub, N.~Hoffmann, and N.~Akhmediev.
\newblock Rogue waves observation in a water wave tank.
\newblock {\em Phys. Rev. Lett.}, 106:204502, 2011.

\bibitem{Chab12}
A.~Chabchoub, N.~Hoffmann, M.~Onorato, and N.~Akhmediev.
\newblock Super rogue waves: observation of a higher-order breather in water
  waves.
\newblock {\em Phys. Rev. X}, 2:011015, 2012.

\bibitem{AkhAnke}
N.~N. Akhmediev and A.~Ankiewicz.
\newblock {\em Solitons, Nonlinear Pulses and Beams}.
\newblock CHAPMAN {\&} HALL, 1997.

\bibitem{Kib10}
B.~Kibler, J.~Fatome, C.~Finot, G.~Millot, F.~Dias, G.~Genty, N.~Akhmediev, and
  J.~M. Dudley.
\newblock The {P}eregrine soliton in nonlinear fibre optics.
\newblock {\em Nature physics}, 6:790--795, 2010.

\bibitem{Kib12}
B.~Kibler, J.~Fatome, C.~Finot, G.~Millot, G.~Genty, B.~Wetzel, N.~Akhmediev,
  F.~Dias, and J.~M. Dudley.
\newblock Observation of {K}uznetsov-{M}a soliton dynamics in optical fibre.
\newblock {\em Scientific Reports}, 2:463, 2012.

\bibitem{DGKM10}
P.~Dubard, P.~Gaillard, C.~Klein, and V.~B. Matveev.
\newblock On multi-rogue waves solutions of the focusing {NLS} equation and
  positon solutions of the {K}d{V} equation.
\newblock {\em Eur. Phys. J. Spec. Top.}, 185:247--261, 2010.

\bibitem{DubMat11}
P.~Dubard and V.~B. Matveev.
\newblock Multi-rogue waves solutions to the focusing {NLS} equation and the
  {KP}-{I} equation.
\newblock {\em Nat.Hazards Earth Syst. Sci}, 11:1--6, 2011.

\bibitem{AKA}
A.~Ankiewicz, D.~J. Kedzora, and N.~Akhmediev.
\newblock Rogue waves triplets.
\newblock {\em Phys. Lett. A}, 375:2782--2785, 2011.

\bibitem{AKA2}
D.~J. Kedzora, A.~Ankiewicz, and N.~Akhmediev.
\newblock Circular rogue wave clusters.
\newblock {\em Phys. Rev. A}, 84:056611, 2011.

\bibitem{Ohta12}
Y.~Ohta and J.~Yang.
\newblock General higer order rogue waves and their dynamics in the nonlinear
  {S}chr{\"o}dinger equation.
\newblock {\em Proc. R. Soc. A}, 468:1716--1740, 2012.

\bibitem{HeF}
J.~S. He, H.~R. Zhang, L.~H. Wang, K.~Porsezian, and A.~S. Fokas.
\newblock A generating mechanism for higer order rogue waves.
\newblock Preprint, arXiv:1209.3742, 2012.
\newblock 5p.

\bibitem{GenDT11}
B.~Guo, L.~Ling, and Q.~P. Liu.
\newblock Nonlinear {S}chr{\"o}dinger equation: Generalized {D}arboux
  transformation and rogue wave solutions.
\newblock {\em Phys. Rev. E}, 85:026607, 2012.

\bibitem{DubMat13}
P.~Dubard and V.~B. Matveev.
\newblock Multi-rogue waves solutions: from the {NLS} equation to the {KP}-{I}
  equation.
\newblock {\em Nonlinearity}, 26:R93--R125, 2013.

\bibitem{GP13}
P~Gaillard.
\newblock Deformations of third-order peregrine breather solutions of the
  nonlinear {S}chr{\"o}dinger equation with four parameters.
\newblock {\em Phys. Rev. E}, 88:042903, 2013.

\bibitem{GP14}
P~Gaillard.
\newblock Ten-parameter deformations of the sixth-order {P}eregrine breather
  solutions of the {NLS} equation.
\newblock {\em Physica Scripta}, 89:015004, 2014.

\bibitem{Aek85e}
N.~Akhmediev, V.~Eleonskii, and N.~Kulagin.
\newblock Generation of periodic trains of picosecond pulses in an optical
  fiber: exact solutions.
\newblock {\em Sov. Phys. JETP}, 62:894--899, 1985.

\bibitem{AK85e}
N.N. Akhmediev and V.I. Korneev.
\newblock Modulation instability and periodic solutions of the nonlinear
  {S}chr{\"o}dinger equation.
\newblock {\em Theor. Math. Phys.}, 69(2):1089--1093, 1986.

\bibitem{IAKe}
G.~L. Alfimov, A.~R. Its, and N.~E. Kulagin.
\newblock Modulation instability of solutions of the nonlinear
  {S}chr{\"o}dinger equation.
\newblock {\em Theor. Math. Phys.}, 84(2):787--793, 1990.

\bibitem{IKe}
A.~R. Its and V.~P. Kotlyarov.
\newblock On a class of solutions of the nonlinear {S}chr{\"o}dinger equation.
\newblock {\em Dokl. Akad. Nauk Ukrain. SSR, Ser. A}, 11:965--968, 1976.
\newblock (Russian).

\bibitem{Chow95}
K.~W. Chow.
\newblock A class of exact, periodic solutions of nonlinear envelope equations.
\newblock {\em J. Math. Phys.}, 36:4125--4137, 1995.

\bibitem{Chow02}
K.~W. Chow.
\newblock A class of doubly periodic waves for nonlinear evolution equations.
\newblock {\em Wave Motion}, 35:71--90, 2002.

\bibitem{Osb00}
A.~R. Osborne, M.~Onorato, and M.~Serio.
\newblock The nonlinear dynamics of rogue waves and holes in deep-water gravity
  wave trains.
\newblock {\em Phys. Lett. A}, 275:386--393, 2000.

\bibitem{Sch02}
A.~Calini and C.~M. Schober.
\newblock Homoclinic chaos increases the likelihood of rogue wave formation.
\newblock {\em Phys. Lett. A}, 298:335--349, 2002.

\bibitem{Sch06}
C.~M. Schober.
\newblock Melnikov analysis and inverse spectral analysis of rogue waves in
  deep water.
\newblock {\em Eur. J. of Mech. B/Fluids}, 25(5):602--620, 2006.

\bibitem{Dys79}
K.~B. Dysthe.
\newblock Note on a modification to the nonliear {S}hr{\"o}dinger equation for
  application to deep water waves.
\newblock {\em Proc. R. Soc. Lond. A}, 369:105--114, 1979.

\bibitem{Dys96}
K.~Trulsen and K.~B. Dysthe.
\newblock A modified nonlinear {S}hr{\"o}dinger equation for broader bandwidth
  gravity waves on deep water.
\newblock {\em Wave Motion}, 24:281--289, 1996.

\bibitem{Dys00}
K.~Trulsen, I.~Kliakhandler, K.~B. Dysthe, and M.~G. Velarde.
\newblock On weakly nonlinear modulation of waves on deep water.
\newblock {\em Phys. Fluids}, 12(10):2433--2437, 2000.

\bibitem{Ind09}
A.~Saini, V.~M. Vyas, S.~N. Pandey, T.~S. Raju, and P.~K. Panigrahi.
\newblock Traveling wave solutions to nonlinear {S}chroedinger equation with
  self-steepening and self-frequency shift.
\newblock Preprint, arXiv:0911.2788, 2009.
\newblock 8p.

\bibitem{Sch05}
A.~L. Islas and C.~M. Schober.
\newblock Predicting rogue waves in random oceanic sea states.
\newblock {\em Phys. of Fluids}, 17(3):031701, 2005.

\bibitem{Ku77e}
E.~A. Kuznetsov.
\newblock Solitons in a parametricaly unstable plasma.
\newblock {\em Sov. Phys., Dokl.}, 22:507--508, 1977.

\bibitem{Ma79}
Ya.~Ch. Ma.
\newblock The pertubed plane-wave solutons of the cubic {S}chr{\"o}dinger
  equation.
\newblock {\em Stud. Appl. Mat.}, 60:43--58, 1979.

\bibitem{Sm12pomie}
A.~O. Smirnov.
\newblock The elliptic breather for the nonlinear {S}chr{\"o}dinger equation.
\newblock {\em J. Math. Sci.}, 192(1):117--125, 2013.

\bibitem{BBEIM}
E.~D. Belokolos, A.~I. Bobenko, V.~Z. Enol'skii, A.~R. Its, and V.~B. Matveev.
\newblock {\em Algebro-geometrical approach to nonlinear evolution equations}.
\newblock Springer Ser. Nonlinear Dynamics. Springer, 1994.

\bibitem{Kalla11}
C.~Kalla.
\newblock Breathers and solitons of generalized nonlinear {S}chr{\"o}dinger
  equation as degenerations of algebro-geometric solutions.
\newblock {\em J. Phys. A}, 44:335210, 2011.

\bibitem{Kalla11a}
C.~Kalla.
\newblock New degeneration of {F}ay's identity and its application to
  integrable system.
\newblock Preprint, arXiv:1104.2568, 2011.
\newblock 40p.

\bibitem{Soleq}
F.~Gesztesy and H.~Holden.
\newblock {\em Soliton equation and their algebro-geometric solutions: Vol. 1,
  (1+1)-dimensional continuous models}, volume~79 of {\em Cambridge Stud. in
  Adv. Math.}
\newblock Cambridge University Press, 2003.

\bibitem{Soleq2}
F.~Gesztesy, H.~Holden, J.~Michor, and G.~Teschl.
\newblock {\em Soliton equation and their algebro-geometric solutions: Vol. 2,
  (1+1)-dimensional discrete models}, volume 114 of {\em Cambridge Stud. in
  Adv. Math.}
\newblock Cambridge University Press, 2008.

\bibitem{Nov74e}
S.~P. Novikov.
\newblock The periodic problem for the {K}orteweg-de {V}ries equation. i.
\newblock {\em Funct. Anal. Appl.}, 8(3):236--246, 1974.

\bibitem{Lax74}
P.~D. Lax.
\newblock Periodic solutions of the {K}-d{V} equations.
\newblock {\em Lect. in Appl. Math.}, 15:85--96, 1974.

\bibitem{DNe}
B.~A. Dubrovin and S.~P. Novikov.
\newblock A periodicity problem for the {K}orteweg-de {V}ries and
  {S}turm-{L}iouville equations. {T}heir connection with algebraic geometry.
\newblock {\em Sov. Math., Dokl.}, 15:1597--1601, 1974.

\bibitem{Mar74e}
V.~A. Marchenko.
\newblock The periodic {K}orteweg-de {V}ries problem.
\newblock {\em Math. USSR Sb.}, 24(3):319--344, 1974.

\bibitem{MvM}
H.~P. McKean and P.~van Moerbeke.
\newblock The spectrum of {H}ill's equation.
\newblock {\em Invent. Math.}, 30:217--274, 1975.

\bibitem{IMe}
A.~R. Its and V.~B. Matveev.
\newblock {S}chr{\"o}dinger operators with finite-gap spectrum and $n$-soliton
  solutions of the {K}d{V} equation.
\newblock {\em Theor. Math. Phys.}, 23(1):343--355, 1975.

\bibitem{DMNe}
B.~A. Dubrovin, V.~B. Matveev, and S.~P. Novikov.
\newblock Nonlinear equations of {K}orteweg-de {V}ries type, finite-band linear
  operators and {A}belian varietes.
\newblock {\em Russ. Math. Surv.}, 31(1):59--146, 1976.

\bibitem{Kr77e}
I.~M. Krichever.
\newblock Methods of algebraic geometry in the theory of non-linear equations.
\newblock {\em Russ. Math. Surv.}, 32(6):185--213, 1977.

\bibitem{Dub81e}
B.~A. Dubrovin.
\newblock Theta functions and non-linear equations.
\newblock {\em Russ. Math. Surv.}, 36(2):11--92, 1981.

\bibitem{Mat08}
V.~B. Matveev.
\newblock 30 years of finite-gap integration theory.
\newblock {\em Phil. Trans. R. Soc. A}, 366:837--875, 2008.

\bibitem{MumII}
D.~Mumford.
\newblock {\em Tata lectures on theta. {II}}, volume~43 of {\em Progress in
  Math.}
\newblock Birkh{\"a}user Boston Inc., Boston, MA, 1984.

\bibitem{Prev85}
E.~Previato.
\newblock Hyperelliptic quasi-periodic and soliton solutions of the nonlinear
  {S}chr{\"o}dinger equation.
\newblock {\em Duke Math. J.}, 52(2):281--545, 1985.

\bibitem{KK12}
C.~Kalla and C.~Klein.
\newblock On the numerical evaluation of algebro-geometric solutions to
  integrable equations.
\newblock {\em Nonlinearity}, 25(3):569--596, 2012.

\bibitem{KK12a}
C.~Kalla and C.~Klein.
\newblock New construction of algebro-geometric solutions to the
  {C}amassa-{H}olm equation and their numerical evaluation.
\newblock {\em Proc. R. Soc. A}, 468(2141):1371--1390, 2012.

\bibitem{Fay}
J.~D. Fay.
\newblock {\em Theta-functions on {R}iemann surfaces}, volume 352 of {\em Lect.
  Notes in Math.}
\newblock Springer, 1973.

\bibitem{Its76e}
A.R. Its.
\newblock Inversion of hyperelliptic integrals and integration of non-linear
  differential equations.
\newblock {\em Vestn. Leningr. Univ. (Mat. Mekh. Astron.)}, 7(2):39--46, 1976.
\newblock (Russian).

\bibitem{Sm12tmfe}
A.~O. Smirnov.
\newblock Solution of a nonlinear {S}chr{\"o}dinger equation in the form of
  two-phase freak waves.
\newblock {\em Theor. Math. Phys.}, 173(1):1403--1416, 2012.

\bibitem{Sm13mze}
A.~O. Smirnov.
\newblock Periodic two-phase ``rogue waves''.
\newblock {\em Math. Notes}, 94(6):897--907, 2013.

\bibitem{Sm94msbe}
A.~O. Smirnov.
\newblock Elliptic solutions of the nonlinear {S}chr{\"o}dinger equation and
  the modified {K}orteweg-de {V}ries equation.
\newblock {\em Russ. Acad. Sci. Sb. Math.}, 82(2):461--470, 1995.

\bibitem{Sm96tmfe}
A.~O. Smirnov.
\newblock The elliptic-in-$t$ solutions of the nonlinear {S}chr{\"o}dinger
  equation.
\newblock {\em Theor. Math. Phys.}, 107(2):568--578, 1996.

\bibitem{Bake}
H.~F. Baker.
\newblock {\em Abel's theorem and the allied theory including the theory of
  theta functions}.
\newblock Cambridge, 1897.

\bibitem{Spre}
G.~Springer.
\newblock {\em Introduction to {R}iemann surfaces}.
\newblock Addison-Wesley, 1957.

\bibitem{FKra}
H.M. Farkas and I.~Kra.
\newblock {\em Riemann surfaces}.
\newblock Springer, New York, 1980.

\bibitem{Sm87msbe}
A.~O. Smirnov.
\newblock A matrix analogue of the {A}ppell theorem and reduction of
  multidimensional {R}iemann theta-functions.
\newblock {\em Math. USSR Sb.}, 61(2):379--388, 1988.

\bibitem{Sm89tmfe}
A.~O. Smirnov.
\newblock Finite-gap solutions of {A}belian {T}oda chain of genus 4 and 5 in
  elliptic functions.
\newblock {\em Theor. Math. Phys.}, 78(1):6--13, 1989.

\bibitem{Akhe}
N.~I. Akhiezer.
\newblock {\em Elements of the theory of elliptic functions}.
\newblock American Mathematical Society, Providence, RI, 1990.
\newblock Translated from the second Russian edition by H. H. McFaden.

\end{thebibliography}

\end{document}